\documentclass[a4paper,10pt]{article}
\usepackage[latin1]{inputenc}
\usepackage{feynmp}
\usepackage[english]{babel}
\usepackage{amssymb}
\usepackage{cite}
\usepackage{mcite}
\usepackage{graphicx}

\newcommand{\fslash}[1]{\hbox{$#1$}\!\!\!\!/\;}

\newcommand{\half}{\frac{1}{2}}
\newcommand{\dF}{d_\mathrm{F}}
\newcommand{\dA}{d_\mathrm{A}}
\newcommand{\CF}{C_\mathrm{F}}
\newcommand{\CA}{C_\mathrm{A}}
\newcommand{\TF}{T_\mathrm{F}}

\newcommand{\mD}{m_\mathrm{D}}
\newcommand{\nF}{\frac{n_\mathrm{F}}{2}}
\newcommand{\NF}{n_\mathrm{F}}
\newcommand{\nS}{n_\mathrm{S}}
\newcommand{\Nc}{N_\mathrm{c}}

\newcommand{\pE}{p_\mathrm{E}}
\newcommand{\pM}{p_\mathrm{M}}
\newcommand{\mE}{m_\mathrm{E}}
\newcommand{\mM}{m_\mathrm{M}}
\newcommand{\dd}{\mathrm{d}}

\newcommand{\Z}{\mathcal{Z}}

\newcommand{\AIC}{\mathcal{I}_C}
\newcommand{\AIE}{\mathcal{I}_E}

\newcommand{\Iint}{\mathcal{I}}
\newcommand{\Mint}{\mathcal{M}}
\newcommand{\Nint}{\mathcal{N}}
\newcommand{\sumint}[1]{\hbox{$\sum$}\!\!\!\!\!\!\!\int_{#1}}

\begin{document}

\begin{titlepage}
\begin{flushright}
HIP-2005-46/TH\\
hep-ph/0510375
\end{flushright}
\begin{centering}
\vfill
 
{\Large{\bf Pressure of the Standard Model at High Temperatures}}

\vspace{0.8cm}

A.~Gynther\footnote{Antti.Gynther@helsinki.fi},
M.~Veps\"al\"ainen\footnote{Mikko.T.Vepsalainen@helsinki.fi}

\vspace{0.8cm}

\vspace{0.3cm}

\emph{
Theoretical Physics Division, 
Department of Physical Sciences,\\
P.O.Box~64, FIN-00014 University of Helsinki, Finland\\}

\vspace*{0.8cm}
 
\end{centering}

\begin{abstract}
We compute the pressure of the standard model at high temperatures in the symmetric phase to three loops, or to $\mathcal{O}(g^5)$ in all coupling constants. We find that the terms of the perturbative expansion in the SU(2) + Higgs
sector decrease monotonically with increasing order, but the large values of the strong coupling constant 
$g_s$ and the Yukawa coupling of the top quark $g_Y$ make the expansion in the full theory converge more slowly.
The final result is observed to be about 10\% smaller than the ideal gas pressure commonly used in cosmological calculations.
\end{abstract}

\vfill
\noindent
 
\vspace*{1cm}

\vfill

\end{titlepage}
\bibliographystyle{h-physrev4}


\section{Introduction}

The most fundamental quantity for the thermodynamics of a finite temperature gauge theory is its free energy. Its partial derivatives give measurable values of thermodynamic quantities and its singularities indicate possible phase transitions. Extensive work has been devoted to the study of the free energy ($F=fV=-pV$) of QCD and electroweak theory in various temperature ranges and with various levels of rigor.

For QCD and low $T$, ($T<T_c\sim\Lambda_\mathrm{QCD}$) one can use hadron gas models, the region around $T=T_c$ can and has to be studied with first principle lattice Monte Carlo methods \cite{Laermann:2003cv,Karsch:2001vs} and at large $T$ the method with controllable
accuracy is the effective theory one \cite{Ginsparg:1980ef,*Appelquist:1981vg,Braaten:1996jr}. This is based on asymptotic freedom and on separating the relevant scales $\pi T$, the electric scale $\mE=gT$ and the magnetic scale $\mM=g^2T$. The computation of the coefficients of the expansion in $g$ has a long history: the results of orders $g^2$ \cite{Shuryak:1977ut,*Chin:1978gj}, $g^3$ \cite{Kapusta:1979fh}, $g^4 \ln g$ \cite{Toimela:1982hv}, $g^4$ \cite{Arnold:1994ps,*Arnold:1995eb}, $g^5$
\cite{Zhai:1995ac,Braaten:1996jr} and $g^6 \ln g$ \cite{Kajantie:2002wa,Kajantie:2003ax} are known. The calculation has also been generalized to the case when the chemical potentials associated with quarks are nonzero \cite{Vuorinen:2003fs}. This expansion has
several interesting finite-$T$-effects: odd powers of $g$ appear, the logarithmic terms are logs of the ratios of the matching scales, $\log(T/\mE)$ and $\log(m_E/\mM)$, and, finally, the coefficient of
the $g^6$ term is not perturbatively calculable \cite{Linde:1980ts,*Gross:1981br}, since vacuum diagrams of all orders in
the loop expansion contribute to it. This is due to the fact that
the effective theory of the magnetic sector is confining. However,
the coefficient of the $g^6$ term can be determined by a combination
of numerical and several involved analytic computations
\cite{Hietanen:2004ew}. One then has the pressure as a well
defined expansion in the coupling constant $g$ determined in the
$\overline{\textrm{MS}}$ scheme.

While both QCD and the electroweak sector of the standard model are gauge field theories, there is
a crucial difference between them: the Landau pole of QCD corresponds to a length scale $1/\Lambda_\mathrm{QCD} \approx 10^{-15}$~m,
the length scale of a nucleon, while for the electroweak theory $1/\Lambda_\mathrm{EW}\approx 10^6$~m, comparable
to the radius of the Earth. Thus, while confinement effects are important in the QCD case near $T_c$, rendering perturbative
calculations unreliable, they are negligible
in the electroweak case and therefore it is, at least in principle, possible to apply perturbative methods down to $T_c$
and even below that. Due to this there has been an extensive amount of work devoted to calculating the properties of the
electroweak phase transition using, for example, perturbative 1-loop \cite{Anderson:1991zb,Carrington:1991hz,Dine:1992wr,Kapusta:1990qc} 
and 2-loop \cite{Arnold:1992rz,*Arnold:1992rz:err,Farakos:1994kx,Fodor:1994bs} effective potential calculations. Those methods are reliable only for small
Higgs masses and the complete solution of the problem required first a perturbative matching of the full 4-dimensional
theory to an effective 3-dimensional theory \cite{Kajantie:1995dw} and then numerically solving the phase diagram from
the effective theory using lattice Monte Carlo techniques 
\cite{Kajantie:1996mn,Kajantie:1995kf,Kajantie:1996qd,Karsch:1996yh,Gurtler:1997hr}. The phase
diagram was observed to have a first order line which ends in a 2nd order critical point of Ising universality class 
\cite{Rummukainen:1998as}. Similar techniques have been used to solve the phase diagram also when the external U($1$) magnetic
field \cite{Kajantie:1998rz} or the chemical potentials related to the baryon and lepton numbers 
\cite{Gynther:2003za} are nonzero. The phase diagram has also been solved with numerical studies of the full 4-dimensional theory \cite{Csikor:1998eu}. Grand unified theories have been studied in \cite{Toimela:1983qf,Rajantie:1997pr}.

The structure of the phase diagram and the properties of the phase
transition (latent heat, interface tension, correlation lengths,
order parameter discontinuities) only depend on the discontinuities
of the free energy, not on its value. In the effective theory approach
there is thus an important theoretical step missing, the matching
of the value of the free energy itself. For QCD this problem was
formulated and solved in \cite{Braaten:1996jr,Kajantie:2002wa}.
The purpose of this paper is to do the same for the standard model.
At the same time we obtain the pressure to order $g^5$.

To this end we have to carry out a number of rather extensive
computations. First, we compute the 3-loop free energy in the full 4d standard model by evaluating the
3-loop finite-$T$ sum-integrals in the $\overline{\textrm{MS}}$ scheme. Schematically,
we need $F=1+g^2+g^4(1/\epsilon+1)$, where the $g^4$ coefficient contains $1/\epsilon$ terms due to 
infrared (IR) divergences which then cancel against the ultraviolet (UV) divergences of the effective theory.
The UV divergences of the full theory are cancelled by the standard renormalization
procedure. Second, we determine the 2-loop screening masses of the gauge fields $A_0$, $B_0$ 
in the full 4d theory in the $\overline{\textrm{MS}}$ scheme. Schematically, we need
$m^2 = g^2(1+\epsilon)+g^4$. We then repeat the same for the fundamental scalar mass, which
is present already on the tree-level but gets thermal corrections. Unlike the screening masses,
it has additional divergences, so we need $m^2 = -\nu^2 + g^2(1+\epsilon)+g^4(1/\epsilon+1)$. 
Finally, we compute the 3-loop free energy in the 3d effective theory.
Schematically, we need $f=m^3+g_3^2m^2+g_3^4 m$.

Our final result will have some qualitative differences to the pressure of hot QCD. There is a mass scale $\nu$
independent of $T$ in the Lagrangian which gives rise to terms such as $\nu^2T^2$, not present in $p_\mathrm{QCD}$.
Another difference is related to renormalization of the fundamental scalar mass in 3d effective theory. 
In a gauge theory with no scalar particles all the parameters of 
the 3d theory are finite to order $g^4$, whereas in this theory the mass of the fundamental scalar
contains divergences requiring renormalization. We also have terms of order $g^5\ln\, g$, which
cancel in QCD. In addition, the different mass scales
$\mD$, $\mD'$ and $m_3$ lead to terms of type $g^5 \ln m_1/m_2$, where $m_i$ are some combinations
of these masses.

The paper is organized as follows: in section \ref{sec:setting} we explicitly define the theory we are working with
and fix various conventions. In section \ref{sec:separation} we briefly review the method of dimensional reduction. Sections
\ref{sec:4dpres}, \ref{sec:3dparams} and \ref{sec:3dpres} contain the calculations. Finally, in section \ref{sec:numerics}
we discuss the result, the details of which are given in Appendices \ref{app:pEparams} -- \ref{app:diagrams}.


\section{The basic setting}
\label{sec:setting}

The theory we consider is the $\textrm{SU}(3)_c \times \textrm{SU}(2)_L \times \textrm{U}(1)_Y$ standard model with $\NF=3$ families of fermions and $\nS=1$ fundamental scalar doublets, and the quantity we will evaluate is the pressure of this theory at high temperatures. The theory is specified by the Euclidean action (in the units $\hbar = c = 1$)
\begin{eqnarray}
S & = & \int_0^\beta\mathrm{d}\tau\int\mathrm{d}^d x \mathcal{L} \\  
\mathcal{L} & = & \frac{1}{4}G_{\mu\nu}^a G_{\mu\nu}^a + \frac{1}{4}F_{\mu\nu} F_{\mu\nu} + \frac{1}{4}W_{\mu\nu}^a W_{\mu\nu}^a  
+ D_\mu\Phi^\dagger D_\mu\Phi - \nu^2\Phi^\dagger\Phi + \lambda(\Phi^\dagger\Phi)^2 \nonumber \\
& & + \bar{l}_L\fslash{D}l_L + \bar{e}_R\fslash{D}e_R 
+ \bar{q}_L\fslash{D}q_L + \bar{u}_R\fslash{D}u_R + \bar{d}_R\fslash{D}d_R + ig_Y\left(\bar{q}_L\tau^2\Phi^\ast t_R - \bar{t}_R(\Phi^\ast)^\dagger\tau^2q_L\right), 
\end{eqnarray}
where
\begin{eqnarray}
G_{\mu\nu}^a & = & \partial_\mu A_\nu^a - \partial_\nu A_\mu^a + g\epsilon^{abc}A_\mu^b A_\nu^c, \quad \quad 
F_{\mu\nu} \;\;\; = \;\;\; \partial_\mu B_\nu - \partial_\nu B_\mu, \nonumber \\
W_{\mu\nu}^a & = & \partial_\mu C_\nu^a - \partial_\nu C_\mu^a + g_sf^{abc}C_\mu^b C_\nu^c \nonumber \\
D_\mu \Phi & = & \partial_\mu \Phi - \frac{ig}{2}A_\mu^a \tau^a \Phi  + \frac{ig'}{2}B_\mu \Phi,\quad \quad
D_\mu \Phi^\dagger \,\,\, = \,\,\, (D_\mu \Phi)^\dagger, \nonumber \\
\fslash{D}l_L & = & \gamma_\mu\left(\partial_\mu l_L - \frac{ig}{2}A_\mu^a\tau^al_L + \frac{ig'}{2}B_\mu l_L \right), \nonumber \\
\fslash{D}e_R & = & \gamma_\mu\left(\partial_\mu e_R + ig'B_\mu e_R\right) \nonumber \\
\fslash{D}q_L & = & \gamma_\mu\left(\partial_\mu q_L - \frac{ig}{2}A_\mu^a\tau^a q_L - \frac{ig'}{6}B_\mu q_L - ig_s C_\mu^a T^a q_L\right), \nonumber \\
\fslash{D}u_R & = & \gamma_\mu\left(\partial_\mu u_R - \frac{2ig'}{3}B_\mu u_R - ig_s C_\mu^a T^a u_R \right) \nonumber \\
\fslash{D}d_R & = & \gamma_\mu\left(\partial_\mu d_R + \frac{ig'}{3}B_\mu d_R - ig_s C_\mu^a T^a d_R \right).
\end{eqnarray}
Here $A_\mu^a$, $B_\mu$ and $C_\mu^a$ are gauge bosons of weak-, hyper- and strong interactions, respectively; $\Phi$ is the fundamental scalar doublet; $l_L$ and $e_R$ are the left-handed lepton doublets and the right-handed lepton singlets (wrt. weak charge), and $q_L$, $u_R$ and $d_R$ are the left-handed quark doublets and the right-handed up and down -type quark singlets. Only the Yukawa coupling for the top quark is taken into account. Summation over different families is assumed. Also, $d = 3-2\epsilon$ in dimensional regularization, $\mu,\;\nu = 0,...,d$. The gamma matrices are defined in Euclidean space so that $\{\gamma_\mu,\gamma_\nu\} = 2\delta_{\mu\nu}$, $\{\gamma_5,\gamma_\mu\}=0$ and $\textrm{Tr}\, \gamma_5 \gamma_\mu \gamma_\nu \gamma_\rho \gamma_\sigma \propto \epsilon_{\mu\nu\rho\sigma}$. The color indices are $a = 1,...,\dA$ for the weak interaction and $a = 1,...,\Nc^2-1$ for the strong interaction. The different group theory factors for SU($N$) with generators $T^a$ are defined as:

\parbox{0.51\textwidth}{
\begin{eqnarray*}
\TF\delta^{ab} & = & \mathrm{Tr}\;T^a T^b,\\ 
\CA\delta^{ab} & = & f^{ace}f^{bce},
\end{eqnarray*}}
\parbox{0.46\textwidth}{
\begin{eqnarray}
\CF\delta_{ij} & = & \left[T^a T^a\right]_{ij}, \\
\dA & = & \delta^{aa},\quad \dF \; = \; \delta_{ii}.
\end{eqnarray}}
For SU(2) with $T^a = \tau^a/2$ they are $\TF = 1/2$, $\CF = 3/4$, $\CA = 2$, $\dA = 3$ and $\dF = 2$.

The momentum integrations are done using dimensional regularization for both IR and UV divergences. The dimensionful parameter is chosen according to the $\overline{\textrm{MS}}$ scheme, which amounts to replacing the scale parameter $\mu$ by
\begin{equation}
	\Lambda = \mu \left(\frac{e^\gamma}{4\pi}\right)^{-1/2}.
\end{equation}
All couplings are implicitly scaled to their 4d ($\epsilon=0$) dimension with $\mu$, so that e.g.~$g^2=\mu^{-2\epsilon}\hat{g}^2$, where $\hat{g}$ is the coupling in the $4-2\epsilon$-dimensional Lagrangian, $[\hat{g}]=\epsilon$. We use the Feynman gauge ($\xi=1$) for the gauge particle propagators at all stages of the calculation. The final result should, of course, be gauge independent, since pressure is a physical quantity, but we have not checked this explicitly.

The theory contains six couplings that run with the renormalization scale: gauge couplings $g'$, $g$ and $g_s$, the fundamental scalar quartic self-coupling $\lambda$ and its mass parameter $\nu^2$, and $g_Y$. The counterterms in the standard model can be found in \cite{Arnold:1992rz,*Arnold:1992rz:err}. However, the terms proportional to $\lambda$ were neglected there, while the terms including $g'^2$ were dropped in \cite{Kajantie:1995dw}, so for completeness we list the running of the needed parameters here:
\begin{eqnarray}
	\nu^2(\Lambda) &=& \nu^2(\mu) +\frac{1}{8\pi^2}\left( -3\CF g^2 -3\frac{1}{4}g'^2 +N_c g_Y^2 +2(\dF+1)\lambda \right)
		\nu^2 \ln\frac{\Lambda}{\mu}, \label{eq:nurunning} \\
	\lambda(\Lambda) &=& \lambda(\mu) + \frac{1}{8\pi^2}\left( \frac{3}{\dF+1}\left(\CF^2+\CF\TF-\frac{1}{4}\CA\CF\right)g^4 +\frac{3}{16}g'^4 +\frac{3}{2}\frac{\CF}{\dF+1}g^2 g'^2 \right. \nonumber \\
& & \left. \hspace{2cm} -6\CF\lambda g^2 -\frac{3}{2}\lambda g'^2 +(8+2\dF) \lambda^2 -3g_Y^4 +2\Nc \lambda g_Y^2 \right) \ln\frac{\Lambda}{\mu}, \label{eq:lambdarunning} \\
	g_Y^2(\Lambda) &=& g_Y^2(\mu) +\frac{1}{8\pi^2}\left[  \left(\frac{3}{2}+\Nc\right)g_Y^2 -\frac{3}{\Nc}\left(\Nc^2-1\right)g_s^2 -\frac{9}{4}g^2 -\frac{17}{12}g'^2 \right] g_Y^2 \ln\frac{\Lambda}{\mu}, \label{eq:gyrunning} \\
	g^2(\Lambda) &=& g^2(\mu) +\frac{1}{8\pi^2}\left( -\frac{11}{3}\CA +\frac{4}{3}(N_c+1)\nF\TF +\frac{1}{3}\TF\nS\right) g^4 \ln\frac{\Lambda}{\mu}, \label{eq:grunning} \\
	g'^2(\Lambda) &=& g'^2(\mu) +\frac{1}{8\pi^2}\left\{ \left[\frac{2}{3}\left(1+\frac{5}{9}\Nc\right)+\frac{\dF}{6}\left(1+\frac{\Nc}{9}\right)\right]\NF +\frac{1}{3}\frac{\dF}{4}\nS \right\}g'^4\ln\frac{\Lambda}{\mu}, \label{eq:gprunning} \\
	g_s^2(\Lambda) & = & g_s^2(\mu) + \frac{1}{8\pi^2}\left( -\frac{11}{3}\Nc + \frac{4}{3}\NF\right)g_s^4(\mu)\ln\frac{\Lambda}{\mu}. \label{eq:gsrunning}
\end{eqnarray}
Note that $\lambda$ as we have defined it differs from \cite{Arnold:1992rz} by a factor 6. Numerically, we fix the values of these couplings at the scale $\mu=m_Z$ according to their tree-level relation to different physical parameters:

\parbox{0.45\textwidth}{
\begin{eqnarray*}
\nu^2(m_Z) & = & \frac{1}{2}m_H^2, \\
g_Y^2(m_Z) & = & 2\sqrt{2}G_\mu m_t^2, \\
g'^2(m_Z) & = & 4\sqrt{2}G_\mu\left(m_Z^2-m_W^2\right),
\end{eqnarray*}}
\parbox{0.52\textwidth}{
\begin{eqnarray}
\lambda(m_Z) & = & \frac{1}{\sqrt{2}}G_\mu m_H^2, \label{eq:numvalues_start} \\
g^2(m_Z) & = & 4\sqrt{2}G_\mu m_W^2, \\
\alpha_s(m_Z) & = & 0.1187, \label{eq:numvalues_stop}
\end{eqnarray}}
where $m_H$ is the unknown mass of the Higgs boson, $m_W=80.43$~GeV, $m_Z=91.19$~GeV and $m_t=174.3$~GeV are the masses of the W and Z bosons and the top quark, respectively, and $G_\mu=1.664\cdot 10^{-5}\;\mathrm{GeV}^{-2}$ is the Fermi coupling constant. We always assume that $\nS=1$ and $\NF=3$ unless stated otherwise, but use the general form to keep track of different contributions. Note, in particular, that the result will not be valid for $\nS > 1$, since the mixing of scalars is not taken into account. We employ a power counting rule $\lambda~\sim~g'^2~\sim~g_s^2~\sim~g_Y^2~\sim g^2$ and assume the temperature to always be so high that the relation $\nu^2 \lesssim g^2T^2$ applies. 

The physical observable we are studying is the pressure, defined by
\begin{equation}
	p(T) = \lim_{V\to\infty}\frac{T}{V}\ln \int \!\mathcal{D}A\mathcal{D}\psi\mathcal{D}\bar{\psi}\mathcal{D}\Phi
	\exp\left(-S\right).
\label{eq:pressure}
\end{equation}
It is normalized such that the (real part of the) pressure of the symmetric phase vanishes at $T=0$.\footnote{Since the symmetric phase is unstable at $T=0$, the pressure there develops an imaginary part when loop corrections are calculated. The imaginary part can be related to the decay rate of the unstable phase \cite{Weinberg:1987vp}.} The purpose is to calculate the pressure up to, and including, order $g^5(1+\ln g)T^4$, employing the power counting rules above. This amounts to calculating 1-, 2- and 3-loop vacuum diagrams contributing to the pressure. These, together with their symmetry factors, are listed in Appendix \ref{app:diagrams}. Other interesting variables, such as entropy and energy densities $s(T)$ and $\epsilon(T)$, can then be evaluated using standard thermodynamic relations, $s(T) = \partial p/\partial T$, $\epsilon(T) = Ts(T) - p(T)$.


\section{Separation of scales}
\label{sec:separation}
In this Section we will shortly review the rationale of dimensional reduction applied to our case. A more complete treatment can be found, e.g.,~in \cite{Kajantie:1995dw}.
 
A straightforward approach in analytic calculations is to use perturbation theory in evaluating the quantities one is interested in. Since we are working in a temperature range where the gauge couplings, due to asymptotic freedom, are small, one could naively hope this to be a consistent procedure. However, in practice the straightforward expansion in $g^2$ is inhibited by various infrared singularities requiring resummations. This in turn leads to the introduction of many different mass scales.
At high temperature and small coupling the dominant energy scale is the temperature $T$, while the electric and magnetic scales $gT$ and $g^2 T$ are suppressed by powers of $g$. The perturbative result includes logarithms of all these, making it impossible to choose the UV cutoff in such a way that there would not be any large logarithms left. This seems to render perturbation theory unusable.

The solution, as is well known \cite{Ginsparg:1980ef,*Appelquist:1981vg}, is to separate the contributions of different scales into successive effective theories, where all the large scales are integrated out one by one. First, we integrate
\begin{equation} \label{eq:sheavyred}
	p(T) \equiv \pE(T) +\frac{T}{V}\ln \int \!\mathcal{D}A_k \mathcal{D}A_0 \mathcal{D}\Phi \exp\left(-S_\mathrm{E}\right),
\end{equation}
where $S_\mathrm{E}$ contains only the static Matsubara modes of the gauge bosons and of the fundamental scalar (Higgs) field. The contributions of the nonzero Matsubara modes and fermions to the pressure show up as the matching constant $\pE$ (Sec.\ref{sec:4dpres}) and in the parameters of $S_\mathrm{E}$ (Sec.\ref{sec:3dparams}). The spatial (magnetic) gauge field components remain massless, while the temporal component gets a thermal mass $\mD \sim gT$. The theory defined by $S_\mathrm{E}$ can then be viewed as a 3d gauge theory with adjoint and a fundamental scalar fields.

The effective theory thus obtained still contains contributions from two scales, $gT$ and $g^2T$, so one more reduction step is useful. We integrate out the scale $gT$ and are left with
\begin{equation} \label{eq:heavyred}
	p(T) \equiv \pE(T) +\pM(T) +\frac{T}{V}\ln \int \!\mathcal{D}A_k \mathcal{D}\Phi \exp\left(-S_\mathrm{M}\right).
\end{equation}
The precise form of the remaining effective theory $S_\mathrm{M}$ depends on the conditions of the system. If the temperature is much higher than the critical temperature of the system\footnote{Within perturbation theory there is always a first order phase transition in electroweak theory.}, then both the adjoint and the fundamental scalars can be integrated out since both of them are massive, $m\sim gT$, and thus the remaining effective theory contains only the spatial gauge fields. The only mass scale of the theory is then provided by the 3d gauge coupling and is of the order $g^2T$. Consequently, the contribution of this theory to the pressure is of the order $g^6$. However, close to the phase transition the fundamental scalar, which drives the transition, becomes light and we are not allowed to integrate it out at the same time as the adjoint scalars. Then the remaining effective theory contains both the spatial gauge fields and the fundamental scalar field, which now has a mass of the order $g^{3/2}T$. To leading order the contribution from this theory to the pressure is then of the order $g^{9/2}T^4 = g^4\sqrt{g}\;T^4$. 

In the present paper we will just consider the case when the temperature is much higher than the critical temperature, and postpone the study of the case when the system is near the phase transition to a later work. The final result of our calculation can then be written as
\begin{equation}
	p(T) = \pE(T) + \pM(T) + p_\mathrm{QCD}(T) + \mathcal{O}(g^6 T^4) \label{eq:final_result},
\end{equation}
where $p_\mathrm{QCD}$ can be taken from \cite{Braaten:1996jr,Arnold:1994ps,*Arnold:1995eb,Kajantie:2002wa,Zhai:1995ac}. One-loop quark diagrams are included in $\pE$, so they must be subtracted from $p_\mathrm{QCD}$.


\section{Calculation of the pressure $\pE$}
\label{sec:4dpres}

The contribution of the nonzero Matsubara modes and fermions to the pressure given by the matching constant $\pE$ is determined by calculating the path integral in Eq.~(\ref{eq:pressure}) without any resummations (Appendix \ref{ft_diagrams}). The calculation involves two different mass scales, the temperature ($2\pi T$) and the mass of the Higgs field ($\nu^2$). Since we assume the temperature to always be so high that $\nu^2 \lesssim g^2 T^2$, we can expand the scalar propagator in powers of $\nu^2$ and keep only terms up to the desired order (integration over the scale $2\pi T$ is infrared safe and thus $\pE$ must be analytic in $\nu^2$). The general form of $\pE(T)$ can then be written as
\begin{eqnarray} \label{eq:4dgenpres}
\pE(T) & = & T^4\Big[\alpha_{E1} + g^2\alpha_{EA} + g'^2\alpha_{EB}
      + \lambda\alpha_{E\lambda} + g_Y^2\alpha_{EY} \nonumber \\ 
& + & \frac{1}{(4\pi)^2}\Big(g^4\alpha_{EAA} + g'^4\alpha_{EBB} +
      (gg')^2\alpha_{EAB} + \lambda^2\alpha_{E\lambda\lambda} + \lambda g^2 \alpha_{EA\lambda}
           + \lambda g'^2 \alpha_{EB\lambda} \nonumber \\
& + & \; g_Y^4\alpha_{EYY} + (gg_Y)^2\alpha_{EAY} + (g'g_Y)^2\alpha_{EBY}
           + \lambda g_Y^2\alpha_{EY\lambda} \nonumber \\
& + & \; (gg_s)^2\alpha_{EAs} + (g'g_s)^2\alpha_{EBs} + (g_Yg_s)^2\alpha_{EYs} \Big) \Big] \nonumber \\
& + & \nu^2T^2\Big[\alpha_{E\nu} + \frac{1}{(4\pi)^2}\big(g^2\alpha_{EA\nu} + g'^2\alpha_{EB\nu} + \lambda\alpha_{E\lambda\nu}
    + g_Y^2\alpha_{EY\nu}\big)\Big] \nonumber \\
& + & \frac{\nu^4}{(4\pi)^2}\alpha_{E\nu\nu} + T^4\cdot{\cal O}(g^6).
\end{eqnarray}
The values of all the coefficients $\alpha_\mathrm{E}$ can be found in Appendix \ref{app:pEparams}. The contribution coming solely from QCD is not included.

All the couplings in the expression above are the renormalized couplings and thus run according to Eqs.~(\ref{eq:nurunning})--(\ref{eq:gsrunning}). However, not all the $1/\epsilon$ poles are cancelled by
the renormalization procedure as can be explicitly seen in the coefficents $\alpha_\mathrm{E}$. The remaining
poles correspond to the IR divergences and are only cancelled when the contribution from the effective theories
to the pressure is taken into account.


\section{Parameters of the 3d theory $S_\mathrm{E}$}
\label{sec:3dparams}

The 3d effective electroweak theory has the general form
\begin{eqnarray} \label{eq:3daction}
S_\mathrm{E} &=& \int\!\dd^3x\left\{ \frac{1}{4}G_{ij}^a G_{ij}^a +\frac{1}{4}F_{ij}F_{ij} +(D_i\Phi)^\dagger(D_i\Phi)
	+m_3^2\Phi^\dagger\Phi +\lambda_3(\Phi^\dagger\Phi)^2 \right. \nonumber \\
	&&+\half(D_i A_0^a)^2 +\half\mD^2 A_0^a A_0^a +\frac{1}{4}\lambda_A (A_0^a A_0^a)^2 +\half(\partial_i B_0)^2
	+\half\mD'^2 B_0 B_0 \nonumber \\
	&& \left. +h_3\Phi^\dagger\Phi A_0^a A_0^a +h_3'\Phi^\dagger\Phi B_0 B_0
	-\half g_3 g_3' B_0\Phi^\dagger A_0^a \tau^a\Phi \right\},
\end{eqnarray}
where $G_{ij}^a = \partial_i A_j^a - \partial_j A_i^a +g_3 \epsilon^{abc}A_i^b A_j^c$, $F_{ij}=\partial_i B_j -\partial_j B_i$,
$D_i\Phi = (\partial_i-ig_3\tau^a A^a_i/2 +ig_3'B_i/2)\Phi$ and $D_i A_0^a = \partial_i A_0^a +g_3 \epsilon^{abc}A_i^b A_0^c$. All the couplings and masses in (\ref{eq:3daction}) can be determined to the required order in $g^2$ by matching the static
Green's functions computed in both the effective and the original theory.

\subsection{Coupling constants}
\label{ssec:couplings}

The 2-loop diagrams in the 3d theory are of the order $g_3^2 m^2 \sim \mathcal{O}(g^4)$. Therefore, the leading order
results for couplings are enough for our purposes, while the corrections would contribute at $g^6$. At tree-level the reduction to the 3d theory only includes scaling the fields by $\sqrt{T}$, and therefore matching the Green's functions gives
\begin{equation} \label{eq:couplingmatch}
\begin{array}{rclrcl}
	g_3^2 &=& g^2 T, & g_3'^2 &=& g'^2 T, \\
	\lambda_3 &=& \lambda T, & \lambda_A &=& \mathcal{O}(g^4), \\
	h_3 &=& \frac{1}{4}g^2 T, & h_3' &=& \frac{1}{4}g'^2 T.
\end{array}
\end{equation}
The quartic couplings of the adjoint scalars, $\lambda_{A,B}$, are not needed at this order. Expressions for them can be found in \cite{Kajantie:1995dw}. Note that the relations above hold for (dimensionful) parameters in the $d$-dimensional Lagrangian. The dimensional regularization scale does not need to be the same in 4d and 3d, which gives $\mathcal{O}(\epsilon)$ corrections to the above matching formulas, e.g.~$g_3^2 = (\Lambda/\mu_3)^{2\epsilon}g^2 T$, where $\Lambda$ and $\mu_3$ are the 4d and 3d dimensional regularization scales, respectively. At the end of the calculation we are going to set 
$\Lambda = \mu_3$. When properly renormalized, the 4d theory has only IR divergencies left and the scale $\Lambda$ should be interpreted as the factorization scale separating the full and the effective theory.

\subsection{Mass parameters}
\label{ssec:masses}

In general, the mass parameters of the effective theory can be found by comparing the poles of static propagators in both theories. In the full theory we have for the pole of the propagator
\begin{equation}\label{eq:mmatch_full}
	k^2 + m^2 + \Pi(k^2) = k^2 + m^2 + \overline{\Pi}(k^2) +\Pi_3(k^2) = 0,
\end{equation}
at $k^2=\mathbf{k}^2=-m_\mathrm{eff}^2$, $k_0=0$. Here $\Pi_3(k^2)$ is the contribution of $n=0$ modes only, and this part is also correctly produced by the effective theory, where the same propagator reads
\begin{equation}
	k^2 + m_3^2 +\Pi_3(k^2) = 0 \quad \textrm{at} \quad k^2=-m_\mathrm{eff}^2.
\end{equation}

For $m \lesssim gT$ the leading order solution is $k \sim gT$. Since $\overline{\Pi}(k^2)$ has no infrared divergence, we can expand it in $k^2$,
\begin{equation}
	\overline{\Pi}(k^2)=\overline{\Pi}(0)+k^2\frac{\dd}{\dd k^2}\overline{\Pi}(0)+\ldots
\end{equation}
Up to $\mathcal{O}(g^4)$ Eq.~(\ref{eq:mmatch_full}) then reads
\begin{equation}
	k^2\left(1+\frac{\dd}{\dd k^2}\overline{\Pi}^{(1)}(0)\right) +m^2+\overline{\Pi}^{(1)}(0)+\overline{\Pi}^{(2)}(0)
	 +\Pi_3(k^2)=0,
\end{equation}
and the matching condition can be read from
\begin{equation}
	m_3^2+\Pi_3(k^2) = \left(1-\frac{\dd}{\dd k^2}\overline{\Pi}^{(1)}(0)\right)
	\left[ m^2+\overline{\Pi}^{(1)}(0)+\overline{\Pi}^{(2)}(0) +\Pi_3(k^2) \right],
\end{equation}
giving
\begin{equation}
	m_3^2 = m^2 +\overline{\Pi}^{(1)}(0)+\overline{\Pi}^{(2)}(0) -\left(m^2 +\overline{\Pi}^{(1)}(0)\right) 
	\frac{\dd}{\dd k^2}\overline{\Pi}^{(1)}(0).
\end{equation}
Note that $\Pi_3 \sim g^2mT \sim g^3 T^2$ cancels between the two equations, since we only need terms up to $g^4$.
There are $1/\epsilon$-divergencies at $m_3^2 g_3^2$ order in the free energy computed in this theory, so we will also need the $g^2\epsilon$-terms in the masses.


\subsubsection{Adjoint scalar masses}
\label{sssec:adjmass}

Applying the procedure described above to the static $A_0$ and $B_0$ propagators gives, after calculating all the 2-loop corrections to them,
\begin{eqnarray}
	\mD^2 &=& T²\left[ g²\left(\beta_{E1} +\beta_{E2}\epsilon +\mathcal{O}(\epsilon^2)\right) 
	+\frac{g^4}{(4\pi)^2}\left(\beta_{E3}+\mathcal{O}(\epsilon)\right) +\mathcal{O}(g^6) \right. \nonumber \\
	&& \left. +\frac{g^2}{(4\pi)^2}\left(\beta_{E\lambda}\lambda +\beta_{Es}g_s^2 +\beta_{EY}g_Y^2
		+\beta_{E'}g'^2 +\beta_{E\nu}\frac{-\nu^2}{T^2} \right) \right], \label{eq:matchmd}  \\ 
	\mD'^2 &=& T²\left[ g'²\left(\beta'_{E1} +\beta'_{E2}\epsilon +\mathcal{O}(\epsilon^2)\right) 
	+\frac{g'^4}{(4\pi)^2}\left(\beta'_{E3}+\mathcal{O}(\epsilon)\right) +\mathcal{O}(g'^6)\right. \nonumber \\
	&& \left. +\frac{g'^2}{(4\pi)^2}\left(\beta'_{E\lambda}\lambda +\beta'_{Es}g_s^2 +\beta'_{EY}g_Y^2
		+\beta'_{E}g^2 +\beta'_{E\nu}\frac{-\nu^2}{T^2} \right) \right]. \label{eq:matchmdp}
\end{eqnarray}
The coefficients $\beta$ are listed in Appendix \ref{app:matching}.

In the above equation $g$ and $g'$ are the renormalized couplings, which run as in Eqs.~(\ref{eq:grunning}) and (\ref{eq:gprunning}). Substituting these into Eqs.~(\ref{eq:matchmd}) and (\ref{eq:matchmdp}), we note that at $\mathcal{O}(\epsilon^0)$ all the dependence on the dimensional regularization scale $\Lambda$ is cancelled by the running of $g$'s.

There are also some electroweak corrections to the adjoint scalar mass of QCD. In addition to pure QCD terms already given in \cite{Braaten:1996jr} we have
\begin{equation} \label{eq:qcdmdcorrs}
	m_\mathrm{3E}^2 = m_\mathrm{3E}^2 \big|_\mathrm{QCD} - g_s^2 T^2 \TF^\mathrm{(qcd)} \left( 2\CF\dF\nF g^2 +2\frac{11}{36}\NF g'^2 +\dF g_Y^2 \right),
\end{equation}
and these corrections need to be taken into account in the one-loop term $2m_\mathrm{3E}^3/3\pi$ of the QCD pressure.

Note that the adjoint scalar masses are finite, unlike the fundamental scalar mass below. This is a direct consequence of the fact that the adjoint scalars are actually gauge field components, which have no mass renormalizations, and there are no IR divergences in the matching computation. Electroweak theory also includes a fundamental scalar field, whose mass contains $1/\epsilon$ poles at two-loop level renormalization, since it is not protected by the gauge symmetry.	


\subsubsection{Fundamental scalar mass}
\label{sssec:fundmass}

In \cite{Kajantie:1995dw} the 3d mass of the fundamental scalar was calculated using effective potential methods, but only $\mathcal{O}(\epsilon^0)$ terms were given, and $g'^2$ terms were dropped at two-loop level. We calculated this mass using the same methods as for the adjoint scalar masses, and included also the $\mathcal{O}(g^2\epsilon)$ corrections.

Unlike in the previous section, here we get UV divergencies proportional to $T^2$ that are not cancelled by the counterterms of the 4d theory. These are related to the mass renormalization in the 3d theory, since the matching procedure gives the bare mass $m_{3B}^2$. It serves as an additional check to calculate the 2-loop counterterm dircetly in the 3d theory to see that it precisely cancels the $1/\epsilon$-terms found here.

For the divergent part we have, substituting directly the correct numerical values for group theory factors and setting $\nS=1$,
\begin{equation} \label{eq:m3divergence}
	\delta m_3^2 = \frac{T^2}{(4\pi)^2\epsilon}\left( -\frac{81}{64}g^4 +\frac{7}{64}g'^4 +\frac{15}{32}g^2 g'^2
	-\frac{9}{4}\lambda g^2 -\frac{3}{4}\lambda g'^2 +3\lambda^2 \right),
\end{equation}
while the counterterm in the 3d theory reads
\begin{equation} \label{eq:m3divergence3}
	\frac{1}{(4\pi)^2\epsilon}\left( -\frac{39}{64}g_3^4 +\frac{5}{64}g_3'^2 +\frac{15}{32}g_3^2 g_3'^2
	-\frac{9}{4}\lambda_3 g_3^2 -\frac{3}{4}\lambda_3 g_3'^2 +3\lambda_3^2 +\frac{3}{2}h_3^2 -3h_3 g_3^2 +2h_3'^2 \right).
\end{equation}
The divergent part is independent of the gauge parameter $\xi$ in covariant gauges. This was expected, since the gauge choice in the 3d theory should not depend on that of the 4d theory.

The finite part gives the renormalized 3d mass,
\begin{eqnarray} \label{eq:m3renorm}
	m_3^2(\Lambda) &=& -\nu^2\left[1 +\frac{g^2}{(4\pi)^2}\beta_{\nu A} +\frac{g'^2}{(4\pi)^2}\beta_{\nu B} +\frac{\lambda}{(4\pi)^2}\beta_{\nu\lambda} +\frac{g_Y^2}{(4\pi)^2}\beta_{\nu Y} \right] \nonumber \\
	&&{}+T^2\left[ g^2(\beta_{A1}+\beta_{A2}\epsilon) +g'^2(\beta_{B1}+\beta_{B2}\epsilon)
	+\lambda(\beta_{\lambda 1}+\beta_{\lambda 2}\epsilon) +g_Y^2(\beta_{Y1}+\beta_{Y2}\epsilon) \right. \nonumber \\
	&&{}+\frac{g^4}{(4\pi)^2}\beta_{AA} +\frac{g'^4}{(4\pi)^2}\beta_{BB} +\frac{g^2 g'^2}{(4\pi)^2}\beta_{AB}
	+\frac{\lambda g^2}{(4\pi)^2}\beta_{A\lambda} +\frac{\lambda g'^2}{(4\pi)^2}\beta_{B\lambda}
	+\frac{\lambda^2}{(4\pi)^2}\beta_{\lambda\lambda} \nonumber \\
	&&\left. {}+\frac{g^2 g_Y^2}{(4\pi)^2}\beta_{AY} +\frac{g'^2 g_Y^2}{(4\pi)^2}\beta_{BY} 
	+\frac{g_s^2 g_Y^2}{(4\pi)^2}\beta_{sY} +\frac{\lambda g_Y^2}{(4\pi)^2}\beta_{\lambda Y}
	+\frac{g_Y^4}{(4\pi)^2}\beta_{YY} \right].
\end{eqnarray}
All the different coefficients $\beta_{xy}$ are given in Appendix \ref{app:matching}\@.


\section{Calculation of the pressure $\pM$}
\label{sec:3dpres}
Computing all the 3-loop vacuum diagrams given by the action (\ref{eq:3daction}) (Appendix \ref{et_diagrams}) and treating $m_3$ as being of order $gT$ produces
\begin{eqnarray} \label{eq:3dpres}
	\frac{\pM(T)}{T} &=& \frac{1}{4\pi} \dF \nS \left(m_3^2+\delta m_3^2 \right)^{3/2} \left[ \frac{2}{3} +\epsilon\left(\frac{16}{9}+\frac{4}{3}\ln\frac{\mu_3}{2m_3}\right) \right]
	+\frac{1}{4\pi}\left( \frac{1}{3}\dA \mD^3 +\frac{1}{3} \mD'^3 \right) \nonumber \\
	&+&\frac{1}{(4\pi)^2}\left[-\dF (\dF+1)\nS \lambda_3 m_3^2 -\dF\dA\nS h_3 m_3 \mD -\dF\nS h_3' m_3 \mD' \right. \nonumber \\
	&+& \left. \left(\CF g_3^2 +\frac{1}{4}g_3'^2\right)\nS\dF m_3^2 \left( -\frac{1}{2\epsilon} -\frac{3}{2} -2\ln \frac{\mu_3}{2m_3}\right) +\CA\dA g_3^2\mD^2 \left( -\frac{1}{4\epsilon} -\frac{3}{4} 
		- \ln \frac{\mu_3}{2\mD}\right) \right] \nonumber \\
	&+& \frac{1}{(4\pi)^3}\left[ g_3^4 m_3 B_{AAf} +g_3'^4 m_3 B_{BBf} +g_3^2 g_3'^2 m_3 B_{ABf} +g_3^4 \mD B_{AAa} +g_3^2\lambda_3 m_3 B_{A\lambda f} \right. \nonumber \\
	&+& g_3'^2 \lambda_3 m_3 B_{B\lambda f} +\lambda_3^2 m_3 B_{\lambda \lambda f} +h_3^2 m_3 B_{hhf} +h_3^2 \mD B_{hha} +h_3'^2 m_3 B'_{hhf} +h_3'^2 \mD' B'_{hhb}\nonumber \\
	&+& g_3^2 g_3'^2 m_3 2b(m_3) +g_3^2 g_3'^2\mD b(\mD) +g_3^2 g_3'^2 \mD' b(\mD')+\frac{\dF}{4m_3}(\dA h_3\mD +h_3'\mD')^2 \nonumber \\
	&+& \dF^2 m_3^2 \left( \frac{\dA h_3^2}{2\mD} +\frac{h_3'^2}{2\mD'}\right) 
	+g_3^4\CA\CF\dF\frac{1}{3}\left( \frac{m_3^2}{\mD}\ln\frac{\mD+m_3}{m_3} +\frac{\mD^2}{m_3}\ln\frac{\mD+m_3}{\mD} \right)\nonumber \\
	&+& \dF(\dF+1)\lambda_3(\dA h_3\mD +h_3'\mD') +g_3^2 h_3\mD B_{Aha} +g_3'^2 h_3'\mD' B'_{Bhb} +g_3^2 h_3'\mD' B'_{Ahb} \nonumber \\
	&+&\left. g_3'^2 h_3\mD B_{Bha} +g_3^2 h_3 m_3 B_{Ahf} \right].
\end{eqnarray}
Constants $B_x$ and the function $b(x)$ are given in Appendix \ref{app:3dpressure}. Due to divergences
in $\left(m_3^2+\delta m_3^2 \right)^{3/2}$ we have to expand this term in powers of the coupling constants.
In higher order terms it is enough to use the leading order result
\begin{equation}
	m_3^2 \approx m_T^2 \equiv -\nu^2 +T^2\left( \frac{3}{16}g^2 +\frac{1}{16}g'^2 +\half\lambda +\frac{1}{4}g_Y^2 \right)
\end{equation}
for the thermal Higgs field mass.

All the $1/\epsilon$ poles at $\mathcal{O}(g^5)$ cancel, and substituting the running parameters of Eqs.~(\ref{eq:nurunning}--\ref{eq:gsrunning}) the $g^3$ and $g^5$ orders are seen to be independent of the dimensional regularization scale. The poles at $\mathcal{O}(g^4)$ cancel against those in $\pE$ coming from the heavy ($\pi T$) modes.


\section{Numerical results}
\label{sec:numerics}

In this Section we plot the final result given by Eq.~(\ref{eq:final_result}), into which Eqs.~(\ref{eq:4dgenpres}), (\ref{eq:matchmd}), (\ref{eq:matchmdp}), (\ref{eq:m3divergence}), (\ref{eq:m3renorm}) and (\ref{eq:3dpres}) are inserted, for various values of parameters. In particular, we set $m_H=130$~GeV, which is above the experimental lower limit \cite{Eidelman:2004wy}. Note that vacuum stability considerations lead to a slightly higher limit \cite{Sher:1993mf}, but it turns out that the precise value of $m_H$ does not affect the result much.

\subsection{SU(2) + fundamental Higgs}

Analyzing the result obtained is complicated due to large number of different fields and couplings between them in the complete standard model. The total effective number of degrees of freedom of the theory is 106.75, but the contribution from the Higgs field that drives the transition to this number is just 4 (a complex scalar doublet). Its contribution to the pressure can therefore be expected to be small. Also, because the strong coupling constant $g_s$ and the Yukawa coupling of the top quark $g_Y$ are numerically large when compared to the other gauge couplings and to the Higgs self coupling, their contribution to the pressure dominates over the contribution coming from the Higgs sector that is relevant for the phase transition in the system. It is therefore instructive to consider a simpler SU($2$) + Higgs model for which the total number of degrees of freedom is lower and which does not include couplings that are not directly related to the phase transition.
We achieve this simply by putting $g'^2 = g_s^2 = g_Y^2 = 0$ and $\NF = 0$ in the general result. This theory has also been
studied on lattice \cite{Csikor:1996sp}.

\begin{figure}[tb]
\includegraphics[width=0.87\textwidth]{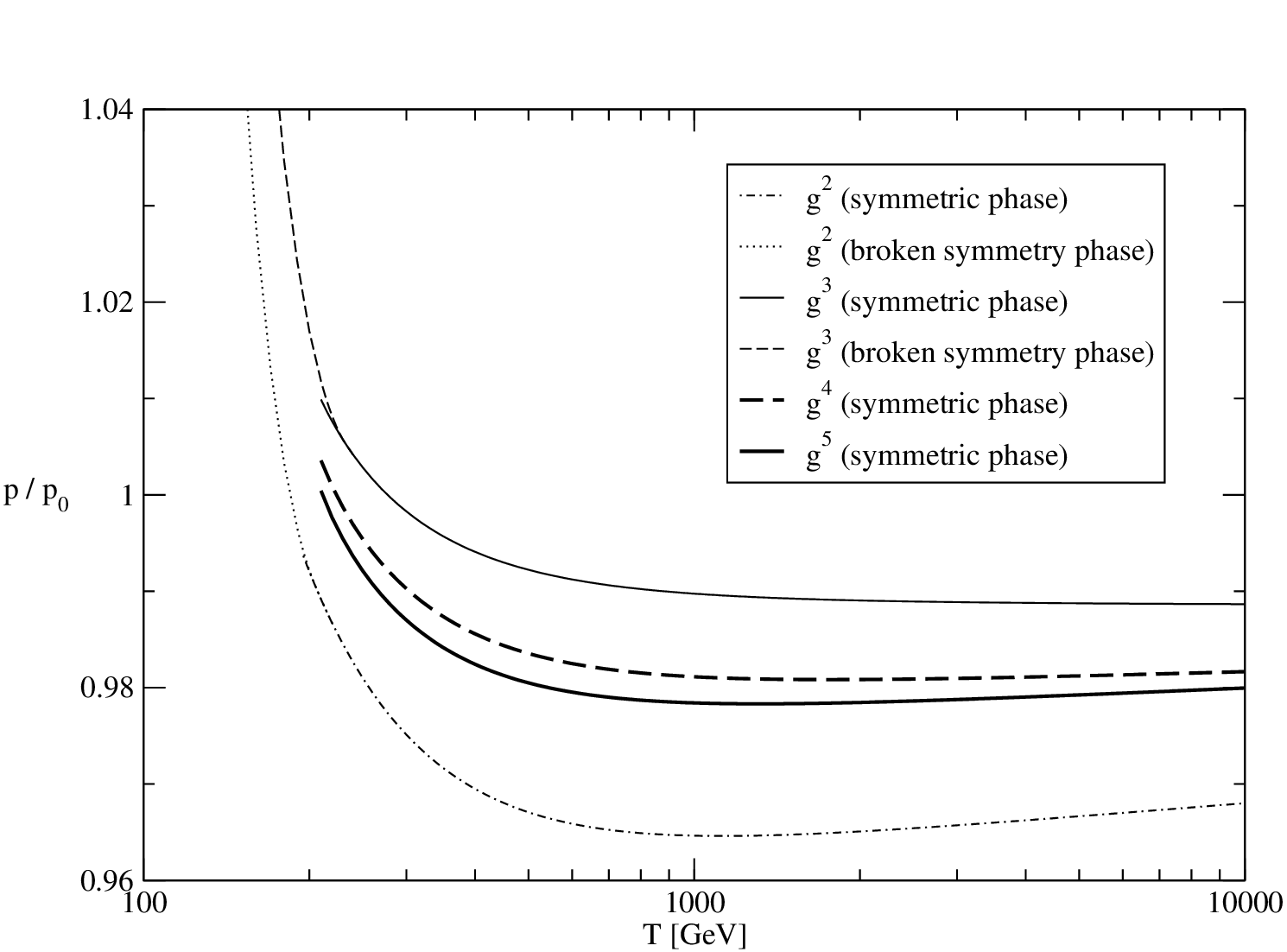}
\caption{The pressure of SU($2$) + fundamental Higgs theory. The Higgs mass is $m_H = 130$~GeV and the W mass is $m_W = 80$~GeV. The critical temperature is $T_c=220$~GeV ($g^3$).}
\label{fig:pressure_su2h}
\end{figure}

To lowest order the pressure of this theory is the ideal gas pressure of SU($2$) gauge bosons and a massless scalar, given by
\begin{equation}
p_0(T) = \frac{\pi^2}{90}T^4\left(2\dA + 2\dF\right) = \frac{\pi^2}{9}T^4.
\end{equation}
Normalizing the results to $p_0$, we plot the pressure of this theory to different orders of the couplings in Fig.~\ref{fig:pressure_su2h}. The mass of the Higgs boson is taken to be $m_H=130$~GeV and the mass of the W boson $m_W=80$~GeV. As can be seen, at high temperatures the introduction of interactions reduces the pressure, but since the coupling is small, this effect is small as well. The perturbative expansion is well behaved in the sense that the absolute
value of each new correction is smaller than that of the previous one. This is in contrast to QCD, where the expansion fluctuates around the ideal gas pressure unless the temperature is taken to be asymptotically large.

The result differs from that of QCD also in the sense that there is another manifest mass scale in the system, the mass
of the Higgs boson. The terms $\nu^2 T^2$ and $\nu^4$ in the expansion of the pressure become more significant and the pressure deviates from the standard Stefan-Boltzmann law $p \sim T^4$ as the temperature gets smaller. Schematically, the pressure of a gas of massive particles is given to leading order by $p \sim T^4(1-g^2) - \sum_i m_i^2 T^2$, where $i$ labels all the particle types in the system and the masses are the thermal masses, $m^2 \sim g^2 T^2$ for the temporal component of the gauge bosons and $m^2 \sim -\nu^2 + g^2T^2$ for the Higgs scalar. Thus, as the temperature is lowered, the pressure, normalized to the ideal gas pressure of massless particles, behaves as $p/p_0 \sim 1 - g^2 + \nu^2/T^2$. This is seen explicitly in Fig.~\ref{fig:pressure_su2h}.

Although the calculation presented in this paper is not, in principle, valid near the phase transition, we have plotted in Fig.~\ref{fig:pressure_su2h} also the pressure of the broken phase to orders $g^2$ and $g^3$, corresponding to a two-loop calculation. This gives us qualitative understanding about the behavior of the pressure near the phase transition.

\begin{figure}[!tb]
\includegraphics[width=0.87\textwidth]{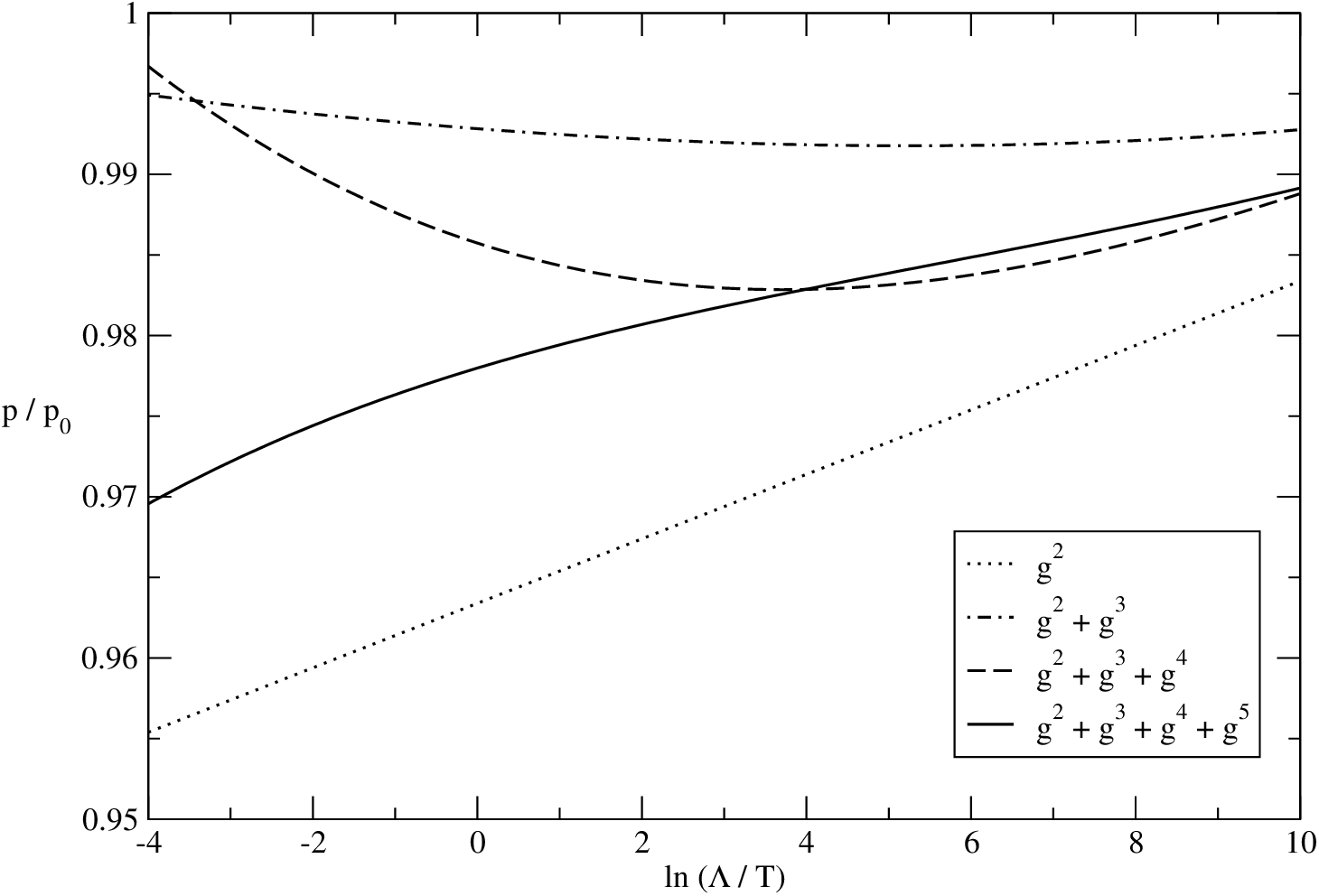}
\caption{The scale dependence of the pressure of SU($2$) + fundamental Higgs theory. The temperature is fixed to $T = 500$~GeV, the Higgs mass is $m_H = 130$~GeV and the W mass is $m_W = 80$~GeV.}
\label{fig:scale_dep_su2h}
\end{figure}

The pressure of the broken phase plotted in Fig.~\ref{fig:pressure_su2h} is given by ($\varphi/\sqrt{2}$ being the
expectation value of the Higgs field) \cite{Arnold:1992rz}
\begin{eqnarray}
p_\mathrm{BP}(T,\varphi) & = & \frac{1}{2}\nu^2\varphi^2 - \frac{1}{4}\lambda\varphi^4 + \frac{\pi^2}{9}T^4 -\frac{13}{192}g^2T^4 -\frac{1}{24}\lambda T^4 \nonumber \\
& & - \frac{T^2}{24}\left(m_H(\varphi)^2 + 3m_{GB}(\varphi)^2 + 9 m_W(\varphi)^2\right) \nonumber \\
& & + \frac{T}{12\pi}\left[\left(m_H(\varphi)^2 + \frac{3}{16}g^2T^2+\frac{1}{2}\lambda T^2\right)^{3/2} + 3\left(m_{GB}(\varphi)^2 + \frac{3}{16}g^2T^2+\frac{1}{2}\lambda T^2\right)^{3/2} \right.\nonumber \\
& & \left.\hspace{1cm} + 6 m_W(\varphi)^3 + 3\left(m_W(\varphi)^2+\frac{5}{6}g^2T^2\right)^{3/2}\right] + \mathcal{O}(g^4),
\end{eqnarray}
where $m_H(\varphi)^2 = 3\lambda\varphi^2 - \nu^2$, $m_{GB}(\varphi)^2 = \lambda\varphi^2 - \nu^2$ and $m_W(\varphi)^2 = 1/4 g^2\varphi^2$ are the zero temperature masses of the particles, and $\varphi = \varphi(T)$ is such that $\partial p_\mathrm{BP} / \partial \varphi^2 = 0$. We can now directly observe that there is a temperature where the pressures of the symmetric phase and of the broken symmetry phase are equal. Below that the pressure of the broken symmetry phase is bigger and thus the symmetry of the theory gets spontanously broken.

Another interesting question is how the scale dependence of the expansion behaves as higher order corrections
are added. If the perturbative expansion is well behaved, one expects the scale dependence to reduce
as more terms are included. This is plotted in Fig.~\ref{fig:scale_dep_su2h}. The temperature is fixed
to $T=500$~GeV. As can be seen, the result depends very weakly on the chosen renormalization scale: varying
the scale within $\Lambda/T \sim 10^{-2}\dots 10^4$ changes the result just about two percent. More specifically,
it is seen that the result up to order $g^3$ is fairly scale independent. This, however, is a numerical
coincidence which stems from the particular values of the parameters $g^2$, $\lambda$ and $\nu$ and
does not appear to have any fundamental reason. The weak scale dependence reappearing in terms of order
$g^4$ and $g^5$ seems to support this conclusion. A similar phenomenon was observed in pure gauge theory in \cite{Arnold:1994ps}, where the scale dependence of the pressure up to order $g^3$ was seen to be much weaker than expected.

\subsection{The standard model}

Here we plot the pressure of the full theory for realistic values of couplings, using tree-level relations between the measured values of $m_W$, $m_Z$, $G_F$, $m_t$, $\alpha_s$ and the parameters in our result, as shown in Eqs.~(\ref{eq:numvalues_start})--(\ref{eq:numvalues_stop}). The unknown Higgs particle mass is set to the lowest experimentally accepted value $m_H=130$~GeV.

\begin{figure}[!t]
\includegraphics[width=0.87\textwidth]{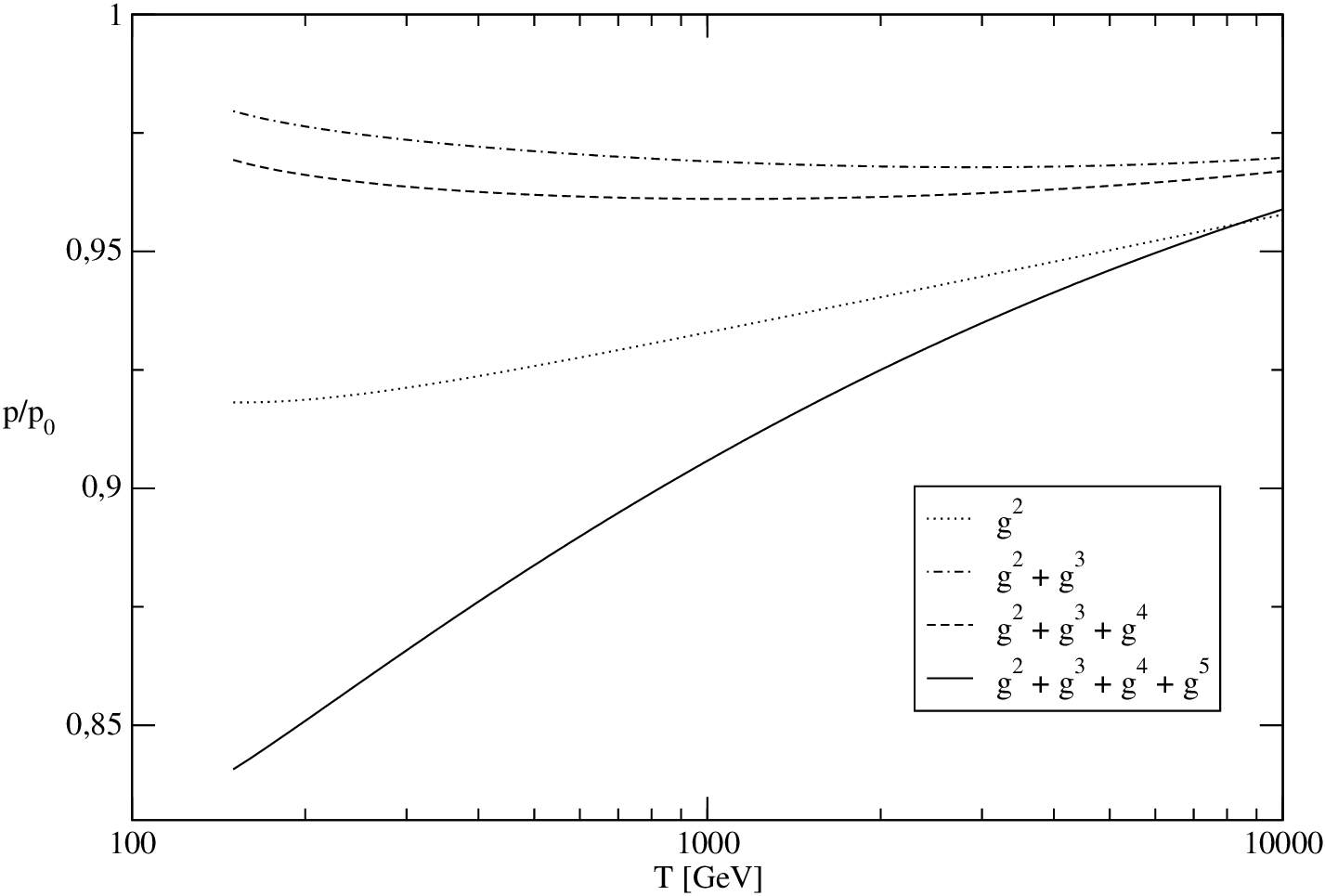}
\caption{Pressure in different orders of perturbation theory.}
\label{fig:orders}
\end{figure}

For an ideal gas of massless SM particles we would have the familiar Stefan--Boltzmann result
\begin{equation}
	p_0 = \frac{\pi^2}{90}T^4 \left( 2 + 2\dA + 2(N_c^2-1) +2\dF +2\frac{7}{8}\NF( \dF +1 +N_c(\dF+2)) \right)
	= 106.75\frac{\pi^2}{90}T^4,
\end{equation}
which actually is $\alpha_{E1}+\textrm{gluons}$. This $T^4$ behavior dominates the pressure, so we again divide by $p_0$ in the plots to see the deviations from massless ideal gas.

The pressure of the full SU(3)$\times$SU(2)$\times$U(1) standard model with 3 families of fermions is plotted in Fig.~\ref{fig:orders}. We show the behavior of the result with increasing orders of perturbation theory to find out the relative size of corrections. Unlike the SU(2) + Higgs case, the result varies strongly with every new order included. This is the known behavior of QCD, and follows from the large values of $g_s$ and $g_Y$, while the higher order terms in $g$, $g'$ and $\lambda$ are small. The $\mathcal{O}(g^5)$ correction is still large enough to push the line downwards near the phase transition. The relative deviation from the ideal gas pressure is of the same order of magnitude as in QCD, which can be explained by the large number of QCD degrees of freedom (79 of the total 106.75). As seen above, in the SU(2) + Higgs theory the deviation is significantly smaller. We have not plotted the pressure all the way down to $T_c$, where the behavior is very singular and the line shoots up to infinity. This stems from the IR divergences in terms like $\mD^2/m_3$, since our assumption $m_3(T) \sim gT$ breaks down near $T_c$ and $m_3$ becomes small.

\begin{figure}[!t]
\includegraphics[width=0.87\textwidth]{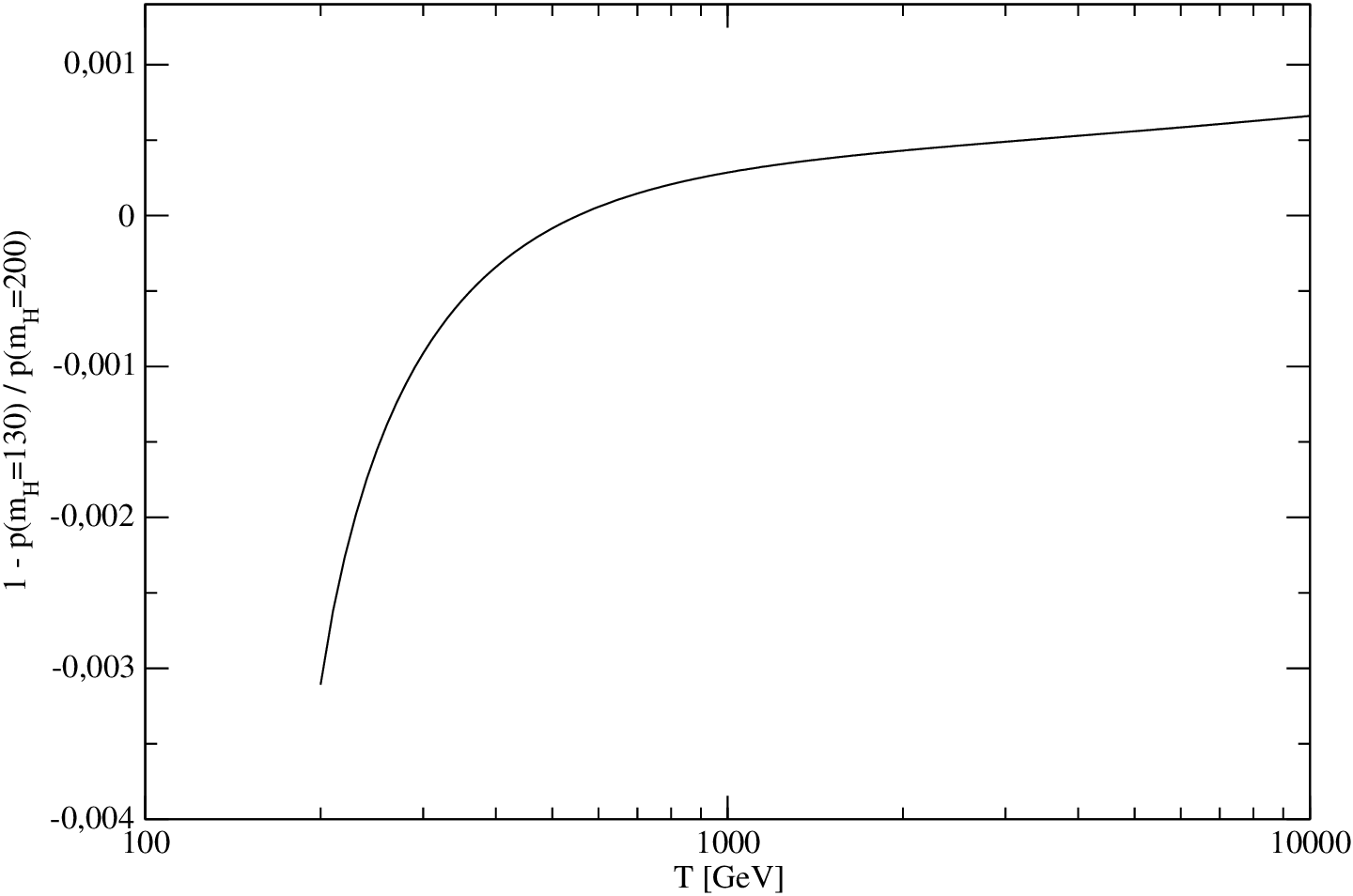}
\caption{Relative difference between pressure at $m_H=130$~GeV and $m_H=200$~GeV.}
\label{fig:higgs_dep}
\end{figure}

The effect of varying $m_H$ is shown in Fig.~\ref{fig:higgs_dep}, where the relative difference between pressure at $m_H=130$~GeV and $m_H=200$~GeV is plotted. The Higgs particle mass affects the behavior of the pressure only very weakly, the only change being a slight and almost constant (times $T^4$) increase in the pressure with increasing $m_H$. This was expected since the fundamental scalar only represents 4 of the $\sim 100$ degrees of freedom.

\section{Conclusions}

In this paper we have calculated the pressure, or the free energy, of the standard model at high temperatures to three loops, or to order $g^5$. The result is about 10\% smaller than the ideal gas pressure. The effective number of degrees of freedom is thus accordingly reduced from the standard value of 106.75 used in cosmological computations. The higher order corrections to the pressure are numerically dominated by contributions coming from the strong coupling constant and the Yukawa coupling of the top quark. Neglecting them shows that the perturbative behavior of the underlying gauge + Higgs theory is good, with the absolute value of each new order in the expansion of the pressure being smaller than that of the previous one. This conclusion is supported by the expansion's small dependence on the renormalization scale. The large numerical values of $g_s$ and $g_Y$ imply that the expansion of the pressure of the complete standard model is not as well behaved. However, even then the convergence of the expansion is better than that of the pressure of pure QCD. 

It is possible to extend the expansion presented in this paper by one more order in powers of the coupling constants by using perturbative calculations, to the order $g^6\ln g$. To evaluate that term would require a 4-loop calculation of the vacuum energy densities of the three-dimensional effective field theory in Eq.~(\ref{eq:3daction}). However, it is impossible to give an unambiguous meaning for this term until the complete $g^6$ term is evaluated as well, a task that already requires the use of nonperturbative methods. The convergence of the perturbative expansion is rather fast so one can expect this term to be numerically small.

The calculation performed is valid when the temperature of the system is much higher than the critical temperature of the system. However, unlike in QCD, where the coupling constant becomes large close to the phase transition and renders the perturbative methods unreliable, in electroweak theory it is possible to extend these calculations also to the temperature region close to the phase transition. This requires a modified effective 3d theory in which the adjoint scalars $A_0$ and $B_0$ are integrated out but the fundamental scalar $\Phi$ is kept, assuming $m_3$ is small compared to $\mD$ and $\mD'$. This is left to a forthcoming work \cite{Gynther:2005xg}.

\section*{Acknowledgements}

We thank K.~Kajantie for useful discussions and M.~Laine, Y.~Schr\"oder and A.~Vuorinen for comments and Academy of Finland,
project 77744, for support. AG was supported by the Graduate School for Particle and Nuclear Physics, GRASPANP and MV by Jenny and Antti Wihuri Foundation.


\appendix


\section{Expansion coefficients for $\pE$}
\label{app:pEparams}

\begin{eqnarray}
	\alpha_{E1} & = & \frac{\pi^2}{45}\left\{1+\dA+\dF\nS+\frac{7}{8}\Big[1+\dF+(2+\dF)\Nc\Big]\NF\right\} \\
	\alpha_{EA} & = & -\frac{1}{144}\left[\CA\dA + \frac{5}{2}\CF\dF\nS + \frac{5}{4}\CF\dF(1+\Nc)\NF\right] \\
	\alpha_{EB} & = & -\frac{5}{576}\left\{\frac{1}{2}\dF\nS + \left[1 + \frac{1}{4}\dF + \left(\frac{5}{9}+\frac{1}{36}\dF\right)\Nc\right]\NF\right\} \\
	\alpha_{E\lambda} & = & -\frac{\dF(\dF+1)}{144}\nS \\
	\alpha_{EY} & = & -\frac{5}{288}\Nc \\
	\alpha_{EAA} & = & \frac{1}{12}\left\{\CA^2\dA\left(\frac{1}{\epsilon} + \frac{97}{18}\ln\frac{\Lambda}{4\pi T} + \frac{29}{15} + \frac{1}{3}\gamma + \frac{55}{9}\frac{\zeta'(-1)}{\zeta(-1)} - \frac{19}{18}\frac{\zeta'(-3)}{\zeta(-3)}\right) \right. \nonumber \\
& & \hspace{0.5cm} + \left[\CA\CF\dF\left(\frac{1}{2\epsilon} + \frac{169}{72}\ln\frac{\Lambda}{4\pi T} + \frac{1121}{1440}-\frac{157}{120}\ln 2 + \frac{1}{3}\gamma + \frac{73}{36}\frac{\zeta'(-1)}{\zeta(-1)}-\frac{1}{72}\frac{\zeta'(-3)}{\zeta(-3)}\right) \right. \nonumber \\
& & \hspace{1.5cm} \left. + \CF^2\dF\left(\frac{35}{32}-\ln 2\right)\right]\left(1+\Nc\right)\NF \nonumber \\
& & \hspace{0.5cm} + \CF\TF\dF\left(\frac{5}{36}\ln\frac{\Lambda}{4\pi T} + \frac{1}{144} - \frac{11}{3}\ln 2 + \frac{1}{12}\gamma + \frac{1}{9}\frac{\zeta'(-1)}{\zeta(-1)} - \frac{1}{18}\frac{\zeta'(-3)}{\zeta(-3)}\right)\left(1+\Nc\right)^2\NF^2 \nonumber \\
& & \hspace{0.5cm} + \CF\TF\dF\left(\frac{25}{72}\frac{\Lambda}{4\pi T} - \frac{83}{16} - \frac{49}{12}\ln 2 + \frac{1}{3}\gamma + \frac{1}{36}\frac{\zeta'(-1)}{\zeta(-1)} - \frac{1}{72}\frac{\zeta'(-3)}{\zeta(-3)}\right)\left(1+\Nc\right)\NF\nS \nonumber \\
& & \hspace{0.5cm} + \left[\CA\CF\dF\left(\frac{1}{\epsilon} + \frac{317}{72}\ln\frac{\Lambda}{4\pi T} + \frac{337}{720} + \frac{2}{3}\gamma + \frac{125}{36}\frac{\zeta'(-1)}{\zeta(-1)} + \frac{19}{72}\frac{\zeta'(-3)}{\zeta(-3)}\right) \right. \nonumber \\
& & \hspace{1.5cm} + \CF^2\dF\left(\frac{3}{2\epsilon} + \frac{19}{2}\ln\frac{\Lambda}{4\pi T} + \frac{881}{120} + \frac{3}{4}\gamma + \frac{23}{2}\frac{\zeta'(-1)}{\zeta(-1)} - \frac{11}{4}\frac{\zeta'(-3)}{\zeta(-3)}\right) \nonumber \\
& & \hspace{1.5cm} \left. \left. + \CF\TF\dF\left(\frac{23}{36}\ln\frac{\Lambda}{4\pi T} - \frac{283}{360} + \frac{1}{3}\gamma + \frac{11}{18}\frac{\zeta'(-1)}{\zeta(-1)} - \frac{11}{36}\frac{\zeta'(-3)}{\zeta(-3)}\right)\right]\nS\right\} \\
	\alpha_{EBB} & = & \frac{1}{128}\left\{\left[\dF\left(\frac{1}{\epsilon} + \frac{19}{3}\ln\frac{\Lambda}{4\pi T} + \frac{881}{180} + \frac{1}{2}\gamma + \frac{23}{3}\frac{\zeta'(-1)}{\zeta(-1)} - \frac{11}{6}\frac{\zeta'(-3)}{\zeta(-3)} \right) \right. \right. \nonumber \\
& & \hspace{0.5cm} \left. + \dF^2\left(\frac{23}{54}\ln\frac{\Lambda}{4\pi T} - \frac{283}{540} + \frac{2}{9}\gamma + \frac{11}{27}\frac{\zeta'(-1)}{\zeta(-1)} - \frac{11}{54}\frac{\zeta'(-3)}{\zeta(-3)}\right)\right]\nS \nonumber \\
& & \hspace{0.5cm} + \dF\left[1+\frac{5}{9}\Nc + \frac{\dF}{4}\left(1+\frac{\Nc}{9}\right)\right] \nonumber \\
& & \hspace{1.2cm} \times \left[\frac{25}{27}\ln\frac{\Lambda}{4\pi T} - \frac{83}{60} - \frac{147}{135}\ln 2 + \frac{8}{9}\gamma + \frac{2}{27}\frac{\zeta'(-1)}{\zeta(-1)} - \frac{1}{27}\frac{\zeta'(-3)}{\zeta(-3)}\right]\NF\nS \nonumber \\
& & \hspace{0.5cm} + \left[1+\frac{17}{81}\Nc+\frac{\dF}{16}\left(1+\frac{\Nc}{81}\right)\right]\left(\frac{35}{3}-\frac{32}{3}\ln 2\right)\NF \nonumber \\
& & \hspace{0.5cm} + \left[\left(1+\frac{5}{9}\Nc\right)^2+\frac{\dF}{2}\left(1+\frac{2}{3}\Nc+\frac{5}{81}\Nc^2\right) + \frac{\dF^2}{16}\left(1+\frac{\Nc}{9}\right)^2\right] \nonumber \\
& & \left. \hspace{1.2cm} \times \left(\frac{40}{27}\ln\frac{\Lambda}{4\pi T} + \frac{2}{27}-\frac{176}{45}\ln 2 + \frac{8}{9}\gamma + \frac{32}{27}\frac{\zeta'(-1)}{\zeta(-1)}-\frac{16}{27}\frac{\zeta'(-3)}{\zeta(-3)}\right)\NF^2\right\} \\
	\alpha_{EAB} & = & \frac{1}{16}\left[\CF\dF\left(\frac{1}{\epsilon} + \frac{19}{3}\ln\frac{\Lambda}{4\pi T} + \frac{881}{180} + \frac{1}{2}\gamma + \frac{23}{3}\frac{\zeta'(-1)}{\zeta(-1)} - \frac{11}{6}\frac{\zeta'(-3)}{\zeta(-3)}\right)\nS \right. \nonumber \\
& & \left. \hspace{0.6cm} 
                         + \CF\dF\left(1+\frac{1}{9}\Nc\right)\left(\frac{35}{48} - \frac{2}{3}\ln 2\right)\NF\right] \\
	\alpha_{E\lambda\lambda} & = & \frac{\dF(\dF+1)}{9}\nS\left[\ln\frac{\Lambda}{4\pi T} + \frac{31}{40} + \frac{1}{4}\gamma
                                       + \frac{3}{2}\frac{\zeta'(-1)}{\zeta(-1)} - \frac{3}{4}\frac{\zeta'(-3)}{\zeta(-3)}
                                       + \frac{1}{4}\dF\left(\ln\frac{\Lambda}{4\pi T} + \gamma\right)\right]\\
	\alpha_{EA\lambda} & = & \frac{\dF(\dF+1)}{36}\CF\left(\frac{3}{\epsilon} + 15\ln\frac{\Lambda}{4\pi T} + 11 + 3\gamma
                                 + 12\frac{\zeta'(-1)}{\zeta(-1)}\right)\nS \\
	\alpha_{EB\lambda} & = & \frac{\dF(\dF+1)}{144}\nS\left(\frac{3}{\epsilon} + 15\ln\frac{\Lambda}{4\pi T} + 11 + 3\gamma
                             + 12\frac{\zeta'(-1)}{\zeta(-1)} \right) \\
	\alpha_{EYY} & = & -\frac{1}{32}\Nc\left[\ln\frac{\Lambda}{4\pi T} - \frac{239}{120}- \frac{11}{5}\ln 2
                           +2\frac{\zeta'(-1)}{\zeta(-1)} - \frac{\zeta'(-3)}{\zeta(-3)} \right.\nonumber \\
& &  \left. \hspace{1.1cm} - \Nc\left(\frac{10}{9}\ln\frac{\Lambda}{4\pi T} + \frac{53}{90} - \frac{106}{45}\ln 2 + \frac{4}{9}\gamma
            +\frac{4}{3}\frac{\zeta'(-1)}{\zeta(-1)} - \frac{2}{3}\frac{\zeta'(-3)}{\zeta(-3)} \right)\right]\\
	\alpha_{EAY} & = & \frac{1}{16}\Nc\left(\frac{1}{\epsilon} + \frac{19}{4}\ln\frac{\Lambda}{4\pi T} + \frac{619}{120} - \frac{13}{4}\ln 2 + \gamma + \frac{7}{2}\frac{\zeta'(-1)}{\zeta(-1)} + \frac{1}{4}\frac{\zeta'(-3)}{\zeta(-3)}\right)\\
	\alpha_{EBY} & = & \frac{1}{48}\Nc\left(\frac{1}{\epsilon} + \frac{131}{36}\ln\frac{\Lambda}{4\pi T} + \frac{6563}{1080}
                         -\frac{41}{20}\ln 2 + \gamma + \frac{23}{18}\frac{\zeta'(-1)}{\zeta(-1)} + \frac{49}{36}\frac{\zeta'(-3)}{\zeta(-3)} \right)\\
	\alpha_{EY\lambda} & = & \frac{1}{6}\Nc\left(\ln\frac{\Lambda}{4\pi T} - \ln 2 + \gamma\right) \\
	\alpha_{EAs} & = & \frac{\CF\dF}{12}\left(\Nc^2-1\right)\NF\left(\frac{35}{32} - \ln 2\right) \\
	\alpha_{EBs} & = & \frac{1}{12}\left(\Nc^2-1\right)\NF\left[\frac{175}{288}-\frac{5}{9}\ln 2 + \frac{\dF}{36}\left(\frac{35}{32}-\ln 2\right)\right] \\
	\alpha_{EYs} & = & -\frac{15}{144}\left(\Nc^2-1\right)\left(\ln\frac{\Lambda}{4\pi T}-\frac{62}{75}-\frac{27}{25}\ln 2 
                         +2\frac{\zeta'(-1)}{\zeta(-1)} - \frac{\zeta'(-3)}{\zeta(-3)}\right) \\
	\alpha_{E\nu} & = & \frac{\dF}{12}\nS \\
	\alpha_{EA\nu} & = & -\frac{\CF\dF}{2}\left(\frac{1}{\epsilon} + 3\ln\frac{\Lambda}{4\pi T} + \frac{5}{3} + \gamma + 2\frac{\zeta'(-1)}{\zeta(-1)}\right)\nS \\
	\alpha_{EB\nu} & = & -\frac{\dF}{8}\left(\frac{1}{\epsilon} + 3\ln\frac{\Lambda}{4\pi T} + \frac{5}{3} +\gamma + 2\frac{\zeta'(-1)}{\zeta(-1)} \right) \nS\\
	\alpha_{E\lambda\nu} & = & -\frac{\dF(\dF+1)}{3}\nS\left(\ln\frac{\Lambda}{4\pi T} + \gamma\right) \\
	\alpha_{EY\nu} & = & -\frac{1}{3}\Nc\left(\ln\frac{\Lambda}{4\pi T} -\ln 2 +\gamma \right)\\
	\alpha_{E\nu\nu} & = & \dF\nS\left(\ln\frac{\nu}{4\pi T} -\frac{3}{4} +\gamma \right)
\end{eqnarray}

The normalization $p(T=0)=0$ in the symmetric phase is taken into account in $\alpha_{E\nu\nu}$.


\section{Matching coefficients}
\label{app:matching}

\subsection{Coefficients for the adjoint scalar masses}

\begin{eqnarray}
	\beta_{E1} &=& \frac{1}{3}\left[ \CA +\nF(N_c+1)\TF +\nS\TF \right] \\
	\beta'_{E1} &=& \frac{1}{3}\left[ \left(\frac{11}{36}N_c+\frac{3}{4}\right)\NF +\frac{\dF}{4}\nS \right] \\
	\beta_{E2} &=& \frac{2}{3}\left[ \CA\left( \frac{\zeta'(-1)}{\zeta(-1)}+\ln\frac{\Lambda}{4\pi T} \right) +\TF\nF(N_c+1) \left(\half-\ln 2+\frac{\zeta'(-1)}{\zeta(-1)}+\ln\frac{\Lambda}{4\pi T} \right) \right. \nonumber \\ &&{}\left. +\TF\nS\left(\half +\frac{\zeta'(-1)}{\zeta(-1)}+\ln\frac{\Lambda}{4\pi T}\right) \right] \\
	\beta'_{E2} &=& \frac{2}{3}\left[ \left(\frac{11}{36}N_c+\frac{3}{4}\right)\NF \left(\half-\ln 2+\frac{\zeta'(-1)}{\zeta(-1)}+\ln\frac{\Lambda}{4\pi T} \right) +\frac{\dF}{4}\nS\left(\half +\frac{\zeta'(-1)}{\zeta(-1)}+\ln\frac{\Lambda}{4\pi T}\right) \right] \\
	\beta_{E3} &=& \CA^2\left(\frac{5}{9}+\frac{22}{9}\gamma+\frac{22}{9}\ln\frac{\Lambda}{4\pi T}\right) +\CA\TF \nF(N_c+1)\left(1-\frac{16}{9}\ln 2+\frac{14}{9}\gamma+\frac{14}{9}\ln\frac{\Lambda}{4\pi T}\right) \nonumber \\
	&&{}+\CA\TF\nS\left(\frac{1}{3}+\frac{20}{9}\gamma+\frac{20}{9}\ln\frac{\Lambda}{4\pi T}\right)
	+ \TF^2\left(\nF(N_c+1)\right)^2 \left(\frac{4}{9} -\frac{16}{9}\ln 2 -\frac{8}{9}\gamma-\frac{8}{9}\ln\frac{\Lambda}{4\pi T}\right) \nonumber \\
	&&{}+\TF^2 \nF(N_c+1)\nS \left(\frac{2}{9} -\frac{16}{9}\ln 2 -\frac{10}{9}\gamma -\frac{10}{9}\ln\frac{\Lambda}{4\pi T}\right)
	+ \TF^2\nS^2\left(-\frac{2}{9}-\frac{2}{9}\gamma-\frac{2}{9}\ln\frac{\Lambda}{4\pi T}\right) \nonumber \\
	&&{}-2\CF\TF \nF(N_c+1) +\CF\TF\nS \\
	\beta'_{E3} &=& \left(\frac{11}{36}N_c+\frac{3}{4}\right)^2\NF^2\left(\frac{4}{9} -\frac{16}{9}\ln 2 -\frac{8}{9}\gamma-\frac{8}{9}\ln\frac{\Lambda}{4\pi T}\right)
	+\frac{\dF^2}{16}\nS^2\left(-\frac{2}{9}-\frac{2}{9}\gamma-\frac{2}{9}\ln\frac{\Lambda}{4\pi T} \right) \nonumber \\
	&&{}+\left(\frac{11}{36}N_c+\frac{3}{4}\right)\frac{\dF}{4}\NF\nS\left(\frac{2}{9} -\frac{16}{9}\ln 2 -\frac{10}{9}\gamma -\frac{10}{9}\ln\frac{\Lambda}{4\pi T}\right) \nonumber \\
	&&{}-2\left(\frac{137}{1296}N_c+\frac{9}{16}\right)\NF +\frac{\dF}{16} \nS
\end{eqnarray}
\parbox{0.42\textwidth}{
\begin{eqnarray*}
	\beta_{E\lambda} &=& \frac{2}{3}\TF(\dF+1)\nS \\
	\beta_{Es} &=& -2C_\mathrm{3F}\TF N_c \nF \\
	\beta_{EY} &=& -\frac{1}{6}N_c\TF \\
	\beta_{E'} &=& -2\TF\left( \frac{N_c}{36}+\frac{1}{4}\right)\nF +\TF\frac{1}{4}\nS \\
	\beta_{E\nu} &=& 4\TF\nS
\end{eqnarray*}}
\parbox{0.57\textwidth}{
\begin{eqnarray}
	\beta'_{E\lambda} &=& \frac{2}{3}\frac{\dF}{4}(\dF+1)\nS \\
	\beta'_{Es} &=& -2C_\mathrm{3F}\frac{11}{36}N_c\NF \\
	\beta'_{EY} &=& -\frac{11\dF}{72}N_c \\
	\beta'_E &=& -2\CF\dF\left( \frac{N_c}{36}+\frac{1}{4}\right)\nF +\CF\frac{\dF}{4}\nS \\
	\beta'_{E\nu} &=& 4 \frac{\dF}{4}\nS
\end{eqnarray}}
Taking into account the different group theoretical factors, the pure gauge and fermionic parts of these results agree with 
\cite{Braaten:1996jr}.

\subsection{Coefficients for the fundamental scalar mass}
\parbox{0.45\textwidth}{
\begin{eqnarray*}
	\beta_{\nu A}&=& 3\CF\left( 2\gamma+2\ln\frac{\Lambda}{4\pi T}\right) \\
	\beta_{\nu\lambda}&=& -2(\dF+1)\left( 2\gamma+2\ln\frac{\Lambda}{4\pi T}\right) \\
	\beta_{A1} &=& \frac{1}{4}\CF \\
	\beta_{B1} &=& \frac{1}{4} \left(\half\right)^2 \\
	\beta_{\lambda 1} &=& \frac{1}{6}(\dF+1) \\
	\beta_{Y1} &=& \frac{1}{12}N_c
\end{eqnarray*}}
\parbox{0.54\textwidth}{
\begin{eqnarray}
	\beta_{\nu B}&=& 3\frac{1}{4}\left( 2\gamma+2\ln\frac{\Lambda}{4\pi T}\right) \\
	\beta_{\nu Y}&=& -N_c\left( 4\ln 2 +2\gamma+2\ln\frac{\Lambda}{4\pi T}\right) \\
	\beta_{A2} &=& \CF\half \left(\frac{2}{3}+\frac{\zeta'(-1)}{\zeta(-1)}+\ln\frac{\Lambda}{4\pi T}\right)  \\
	\beta_{B2} &=& \left(\half\right)^2 \half \left(\frac{2}{3}+\frac{\zeta'(-1)}{\zeta(-1)} +\ln\frac{\Lambda}{4\pi T}\right)\\
	\beta_{\lambda 2} &=& \frac{\dF+1}{3} \left(1+\frac{\zeta'(-1)}{\zeta(-1)} +\ln\frac{\Lambda}{4\pi T}\right) \\
	\beta_{Y2} &=& \frac{N_c}{6} \left(1-\ln 2 +\frac{\zeta'(-1)}{\zeta(-1)} +\ln\frac{\Lambda}{4\pi T}\right)
\end{eqnarray}}

\begin{eqnarray}
	\beta_{AA} &=& \left( -\frac{11}{9} -\frac{5}{2}\frac{\zeta'(-1)}{\zeta(-1)}-\frac{2}{3}\gamma -\frac{19}{6}\ln\frac{\Lambda}{4\pi T}\right)\CA\CF
	+\left( 1+\frac{3}{2}\frac{\zeta'(-1)}{\zeta(-1)}+\frac{3}{2}\gamma +3\ln\frac{\Lambda}{4\pi T}\right)\CF^2 \nonumber \\
	&+&\left( \frac{1}{9}+\frac{2}{3}\ln 2 -\frac{2}{3}\gamma -\frac{2}{3}\ln\frac{\Lambda}{4\pi T}\right)\nF(N_c+1)\CF\TF
	+\frac{1}{4}\left(1+\frac{\zeta'(-1)}{\zeta(-1)}+\ln\frac{\Lambda}{4\pi T}\right)\CF(\dF+1) \nonumber \\
	&+&\left( -\frac{2}{9}-\half\frac{\zeta'(-1)}{\zeta(-1)}-\frac{2}{3}\gamma -\frac{7}{6}\ln\frac{\Lambda}{4\pi T}\right)\CF\TF\nS \\
	\beta_{BB} &=& \left( 1+\frac{3}{2}\frac{\zeta'(-1)}{\zeta(-1)}+\frac{3}{2}\gamma +3\ln\frac{\Lambda}{4\pi T}\right)\frac{1}{16}
	+\left( \frac{1}{9}+\frac{2}{3}\ln 2 -\frac{2}{3}\gamma -\frac{2}{3}\ln\frac{\Lambda}{4\pi T}\right)\frac{1}{4}\NF\left(\frac{11}{36}N_c+\frac{3}{4}\right) \nonumber \\
	&+&\left( -\frac{2}{9}-\half\frac{\zeta'(-1)}{\zeta(-1)}-\frac{2}{3}\gamma -\frac{7}{6}\ln\frac{\Lambda}{4\pi T}\right)\frac{\dF}{16}\nS 
	+\left( \frac{1}{4} +\frac{1}{4}\frac{\zeta'(-1)}{\zeta(-1)} +\frac{1}{4}\ln\frac{\Lambda}{4\pi T}\right)\frac{\dF+1}{4} \\
	\beta_{AB} &=& \left( 2+3\frac{\zeta'(-1)}{\zeta(-1)}+3\gamma +6\ln\frac{\Lambda}{4\pi T}\right)\CF\frac{1}{4}
	+\left( \frac{1}{4} +\frac{1}{4}\frac{\zeta'(-1)}{\zeta(-1)} +\frac{1}{4}\ln\frac{\Lambda}{4\pi T}\right)\frac{\dF+1}{2} \\
	\beta_{A\lambda} &=& \left( -\frac{5}{3} -2\frac{\zeta'(-1)}{\zeta(-1)} -2\ln\frac{\Lambda}{4\pi T}\right)\CF(\dF+1) \\
	\beta_{B\lambda} &=& \left( -\frac{5}{3} -2\frac{\zeta'(-1)}{\zeta(-1)} -2\ln\frac{\Lambda}{4\pi T}\right)\frac{1}{4}(\dF+1) \\
	\beta_{\lambda \lambda} &=& \left( 4 +4\frac{\zeta'(-1)}{\zeta(-1)} +4\ln\frac{\Lambda}{4\pi T}\right)(\dF+1) 
	+\left( -\frac{2}{3} -\frac{2}{3}\gamma -\frac{2}{3}\frac{\zeta'(-1)}{\zeta(-1)} -\frac{4}{3}\ln\frac{\Lambda}{4\pi T}\right)(\dF+1)^2 \\
	\beta_{AY} &=& \left( -\frac{1}{12} -\frac{1}{6}\ln2 +\half\gamma +\half\ln\frac{\Lambda}{4\pi T}\right)\CF N_c \\
	\beta_{BY} &=& \left( -\frac{11}{36} -\frac{55}{54}\ln 2 +\frac{17}{18}\gamma +\frac{17}{18}\ln\frac{\Lambda}{4\pi T} \right) \frac{1}{4}N_c \\
	\beta_{sY} &=& \left( -\half +\frac{8}{3}\ln 2 +\gamma +\ln\frac{\Lambda}{4\pi T}\right)C_\mathrm{3F} N_c \\
	\beta_{\lambda Y} &=& \left( -\frac{1}{3}\ln 2 -\frac{2}{3}\gamma -\frac{2}{3}\ln\frac{\Lambda}{4\pi T}\right) (\dF+1)N_c \\
	\beta_{YY} &=& \frac{3}{4}\gamma +\frac{3}{4} \ln\frac{\Lambda}{4\pi T}
\end{eqnarray}


\section{Expansion coefficients for $\pM$}
\label{app:3dpressure}

\begin{eqnarray} 
	B_{AAf} &=& \CF^2\dF\nS \left( -\frac{3}{4\epsilon} -\frac{35}{4} -\frac{\pi^2}{3} +6\ln 2 -\frac{9}{2}\ln\frac{\mu_3}{2m_3} \right) 
	-\CF\TF\dF \left( \frac{1}{4\epsilon} +\frac{4}{3} -\frac{4}{3}\ln 2 +\frac{3}{2}\ln\frac{\mu_3}{2m_3} \right) \nonumber \\
	&& {}+\CA\CF\dF \left( \frac{3}{4\epsilon} +\frac{19}{24} -3\ln 2 +5\ln\frac{\mu_3}{2m_3} -\half\ln\frac{\mu_3}{2(m_3+\mD)} \right) \\
	B_{BBf} &=& \frac{\dF}{16}\nS\left( -\frac{3}{4\epsilon} -\frac{35}{4} -\frac{\pi^2}{3} +6\ln 2 	-\frac{9}{2}\ln\frac{\mu_3}{2m_3} \right)
	 -\frac{\dF^2}{16}\nS \left( \frac{1}{4\epsilon} +\frac{4}{3} -\frac{4}{3}\ln 2 +\frac{3}{2}\ln\frac{\mu_3}{2m_3} \right) \\
	B_{ABf} &=& \CF\dF\nS\left( -\frac{3}{8\epsilon} -\frac{35}{8} -\frac{\pi^2}{6} +3\ln 2 -\frac{9}{4}\ln\frac{\mu_3}{2m_3} \right) \\
	B_{AAa} &=& \CA\CF\dF\nS \left( -\frac{1}{8\epsilon}-\frac{23}{24} -\frac{1}{4}\ln\frac{\mu_3}{2\mD} -\half\ln\frac{\mu_3}{2(m_3+\mD)} \right) \nonumber \\
	&& {}+\CA^2 \dA \left( -\frac{89}{24} +\frac{11}{6}\ln 2 -\frac{\pi^2}{6} \right) \\
	B_{A\lambda f} &=& \CF\dF\nS \left( -4 +8\ln 2  \right)
	+\CF\dF(\dF+1)\nS\left( \frac{1}{\epsilon} +3 +6\ln\frac{\mu_3}{2m_3} \right) \\
	B_{B\lambda f} &=& \frac{1}{4}\dF(\dF+1)\nS\left(\frac{1}{\epsilon} -1 +8\ln 2 +6\ln\frac{\mu_3}{2m_3} \right) \\
	B_{\lambda \lambda f} &=& \dF (\dF+1)\nS\left( -\frac{1}{\epsilon} -8 +4\ln 2 -6\ln\frac{\mu_3}{2m_3} \right) 
	+\half\dF(\dF+1)^2 \\
	B_{Aha} &=& \CF\dF\dA\nS\left( \frac{1}{2\epsilon} +\frac{3}{2} +2\ln\frac{\mu_3}{2m_3} +\ln\frac{\mu_3}{2\mD} \right) \\
	B'_{Bhb} &=& \frac{1}{4}\dF\nS \left( \frac{1}{2\epsilon} +\frac{3}{2} +2\ln\frac{\mu_3}{2m_3} +\ln\frac{\mu_3}{2\mD'} \right) \\
	B'_{Ahb} &=& \CF\dF\nS \left( \frac{1}{2\epsilon} +\frac{3}{2} +2\ln\frac{\mu_3}{2m_3} +\ln\frac{\mu_3}{2\mD'} \right) \\
	B_{Bha} &=& \frac{1}{4}\dF\dA\nS \left( \frac{1}{2\epsilon} +\frac{3}{2} +2\ln\frac{\mu_3}{2m_3} +\ln\frac{\mu_3}{2\mD} \right) \\
	B_{Ahf} &=& \CA\dA\dF\nS \left( \frac{1}{2\epsilon} +\frac{3}{2} +2\ln\frac{\mu_3}{2\mD} +\ln\frac{\mu_3}{2m_3} \right) \\
	B_{hhf} &=& \dF\dA\nS\left( -\frac{1}{2\epsilon} -4 -2\ln\frac{\mu_3}{2(m_3+\mD)} -\ln\frac{\mu_3}{2m_3} \right) \\
	B_{hha} &=& \dF\dA\nS\left( -\frac{1}{2\epsilon} -4 -2\ln\frac{\mu_3}{2(m_3+\mD)} -\ln\frac{\mu_3}{2\mD} \right) \\
	B'_{hhf} &=& \dF\nS\left( -\frac{1}{2\epsilon} -4 -2\ln\frac{\mu_3}{2(m_3+\mD')} -\ln\frac{\mu_3}{2m_3} \right)  \\
	B'_{hhb} &=& \dF\nS\left( -\frac{1}{2\epsilon} -4 -2\ln\frac{\mu_3}{2(m_3+\mD')} -\ln\frac{\mu_3}{2\mD'} \right) \\
	b(x) &=& \CF\dF\nS\left( -\frac{1}{8\epsilon} -1 -\half\ln\frac{\mu_3}{2m_3+\mD+\mD'} -\frac{1}{4}\ln\frac{\mu_3}{2x} \right)
\end{eqnarray}


\section{Diagrams contributing to the pressure}
\label{app:diagrams}

In this appendix we list all the diagrams required for the computation of the pressure. The notation is as follows:
solid lines represent left-handed fermion doublets and right-handed fermion singlets (thick) or just fermion doublets (thin), dashed lines fundamental scalars, dot-dashed lines SU(2) and U(1) (thick) or just SU(2) (thin) adjoint scalars, wavy lines SU(2) and U(1) (thick) or just SU(2) (thin) gauge bosons and curly lines SU(3) gauge bosons. Dotted lines stand for ghosts.

Defining the integration measures as
\begin{eqnarray}
	\sumint{P} & \equiv & \left( \frac{e^\gamma \Lambda^2}{4\pi}\right)^\epsilon T \!\!\!\! \sum_{p_0 = 2n\pi T} 
		\int \! \frac{\dd^{3-2\epsilon} p}{(2\pi)^{3-2\epsilon}}, \\
	\sumint{\{P\} } & \equiv & \left( \frac{e^\gamma \Lambda^2}{4\pi}\right)^\epsilon 
	T \!\!\!\!\!\!\!\! \sum_{p_0 = (2n+1)\pi T} \int \! \frac{\dd^{3-2\epsilon} p}{(2\pi)^{3-2\epsilon}}, \\
	\int_p & \equiv & \left( \frac{e^\gamma \Lambda^2}{4\pi}\right)^\epsilon 
		\int \! \frac{\dd^{3-2\epsilon} p}{(2\pi)^{3-2\epsilon}},
\end{eqnarray}
the diagrams are given in terms of the following integrals:
\begin{eqnarray}
	\Iint_n & \equiv & \sumint{P}  \frac{1}{(P^2)^n}, \\
	\widetilde{\Iint}_n & \equiv & \sumint{ \{P\} } \frac{1}{(P^2)^n}, \\
	\Mint_{i,j} & \equiv & \sumint{PQR} \frac{1}{P^2 Q^2 [R^2]^i [(P-Q)^2]^j (Q-R)^2 (R-P)^2}, \\
	\widetilde{\Mint}_{i,j} & \equiv & \sumint{\{PQR\} } \frac{1}{P^2 Q^2 [R^2]^i [(P-Q)^2]^j (Q-R)^2 (R-P)^2}, \\
	\Nint_{i,j} & \equiv & \sumint{\{PQ\}R} \frac{1}{P^2 Q^2 [R^2]^i [(P-Q)^2]^j (Q-R)^2 (R-P)^2},
\end{eqnarray}
\begin{eqnarray}
	I_n(m) & \equiv & \int_p \frac{1}{(p^2+m^2)^n}, \\
	J_n(m) & \equiv & \int_{pq} \frac{1}{(p^2+m^2)(q^2+m^2)^n(p-q)^2}, \\
	K_n(m) & \equiv & \int_{pq} \frac{1}{(p^2+m^2)(q^2+m^2)[(p-q)^2]^n}, \\
	M_{i,j}(m) & \equiv & \int_{pqr} \frac{1}{(p^2+m^2)(q^2+m^2)(r^2+m^2)^i [(p-q)^2]^j(q-r)^2(r-p)^2}, \\
	N_{i,j}(m) & \equiv & \int_{pqr} \frac{1}{(p^2+m^2)(q^2+m^2)[(q-r)^2+m^2][(r-p)^2+m^2][r^2]^i[(p-q)^2]^j}, \\
	L_{i,j}(m) & \equiv & \int_{pqr} \frac{1}{(p^2+m^2)[(r-p)^2+m^2]^i (q^2+m^2)^j [(q-r)^2+m^2]r^2(p-q)^2},
\end{eqnarray}
\begin{eqnarray}
	\AIC(m_1,m_2,m_3,m_4,0) & \equiv & 
		\int_{pqr} \frac{1}{(p^2+m_1^2)[(r-p)^2+m_2^2][(q-r)^2+m_3^2](q^2+m_4^2)r^2}, \\
	\AIE(m_1,m_2,m_3,m_4) & \equiv & \int_{pqr} \frac{1}{(p^2+m_1^2)(q^2+m_2^2)[(p-r)^2+m_3^2][(q-r)^2+m_4^2]}.
\end{eqnarray}
The integrals are evaluated in Appendices A and B of \cite{Braaten:1996jr} and in \cite{Rajantie:1996cw}.

\subsection{Diagrams in the full theory}
\label{ft_diagrams}

Here we list the results for the diagrams from the 4d theory.

\begin{fmffile}{ft_diags}

\fmfset{curly_len}{1.7mm}




\def\Ring#1{%
  \parbox{30\unitlength}{
  \begin{fmfgraph}(30,30)
	\fmfi{#1}{fullcircle scaled 1h shifted (.5w,.5h)}
  \end{fmfgraph}}}

\def\DiaEight#1#2{%
  \parbox{60\unitlength}{
  \begin{fmfgraph}(60,30)
	\fmfleft{i}
	\fmfright{o}
	\fmf{#1,right}{i,v,i}
	\fmf{#2,right}{o,v,o}
  \end{fmfgraph}}}

\def\Sunset#1#2{%
  \parbox{30\unitlength}{
  \begin{fmfgraph}(30,30)
	\fmfleft{i}
	\fmfright{o}
	\fmf{#1,right}{i,o,i}
	\fmf{#2}{i,o}
  \end{fmfgraph}}}

\def\FSunset#1#2#3{%
  \parbox{30\unitlength}{
  \begin{fmfgraph}(30,30)
	\fmfleft{i}
	\fmfright{o}
	\fmf{#1,right}{i,o}
        \fmf{#2,right}{o,i}
	\fmf{#3}{i,o}
  \end{fmfgraph}}}

\def\Mersu#1#2#3#4{%
  \parbox{30\unitlength}{
  \begin{fmfgraph}(30,30)
	\fmfipath{p}
	\fmfiset{p}{fullcircle scaled 30 rotated -30 shifted (15,15) }
	\fmfipair{v[]}
	\fmfiset{v1}{point 2length(p)/3 of p}
	\fmfiset{v2}{point 0 of p}
	\fmfiset{v3}{point length(p)/3 of p}
	\fmfiset{v4}{(v1+v2+v3)/3}
	\fmfi{#1}{subpath (0,length(p)/3) of p}
	\fmfi{#1}{subpath (length(p)/3,2length(p)/3) of p}
	\fmfi{#2}{subpath (2length(p)/3,length(p)) of p}
	\fmfi{#4}{v1--v4}
	\fmfi{#4}{v4--v2}
	\fmfi{#3}{v3--v4}
  \end{fmfgraph}}}

\def\FMersu#1#2#3#4{%
  \parbox{30\unitlength}{
  \begin{fmfgraph}(30,30)
	\fmfipath{p}
	\fmfiset{p}{fullcircle scaled 30 rotated -30 shifted (15,15) }
	\fmfipair{v[]}
	\fmfiset{v1}{point 2length(p)/3 of p}
	\fmfiset{v2}{point 0 of p}
	\fmfiset{v3}{point length(p)/3 of p}
	\fmfiset{v4}{(v1+v2+v3)/3}
	\fmfi{#1}{subpath (0,length(p)/3) of p}
	\fmfi{#1}{subpath (length(p)/3,2length(p)/3) of p}
	\fmfi{#2}{subpath (2length(p)/3,length(p)) of p}
	\fmfi{#4}{v4--v1}
	\fmfi{#4}{v2--v4}
	\fmfi{#3}{v3--v4}
  \end{fmfgraph}}}

\def\FFMersu#1#2#3#4{%
  \parbox{30\unitlength}{
  \begin{fmfgraph}(30,30)
	\fmfipath{p}
	\fmfiset{p}{fullcircle scaled 30 rotated -30 shifted (15,15) }
	\fmfipair{v[]}
	\fmfiset{v1}{point 2length(p)/3 of p}
	\fmfiset{v2}{point 0 of p}
	\fmfiset{v3}{point length(p)/3 of p}
	\fmfiset{v4}{(v1+v2+v3)/3}
	\fmfi{#1}{subpath (0,length(p)/3) of p}
	\fmfi{#1}{subpath (length(p)/3,2length(p)/3) of p}
	\fmfi{#2}{subpath (length(p),2length(p)/3) of p}
	\fmfi{#4}{v1--v4}
	\fmfi{#4}{v4--v2}
	\fmfi{#3}{v3--v4}
  \end{fmfgraph}}}

\def\DiaV#1#2#3#4{%
  \parbox{30\unitlength}{
  \begin{fmfgraph}(30,30)
	\fmfipath{p}
	\fmfiset{p}{fullcircle scaled 30 rotated -90 shifted (15,15) }
	\fmfipair{v[]}
	\fmfiset{v1}{point 0 of p}
	\fmfiset{v2}{point 3length(p)/8 of p}
	\fmfiset{v3}{point 5length(p)/8 of p}
	\fmfi{#1}{subpath (0,3length(p)/8) of p}
	\fmfi{#2}{subpath (3length(p)/8,5length(p)/8) of p}
	\fmfi{#1}{subpath (5length(p)/8,length(p)) of p}
	\fmfi{#3}{v1--v3}
	\fmfi{#4}{v2--v1}
  \end{fmfgraph}}}

\def\FDiaV#1#2#3#4{%
  \parbox{30\unitlength}{
  \begin{fmfgraph}(30,30)
	\fmfipath{p}
	\fmfiset{p}{fullcircle scaled 30 rotated 135 shifted (15,15) }
	\fmfipair{v[]}
	\fmfiset{v1}{point 3length(p)/8 of p}
	\fmfiset{v2}{point 3length(p)/4 of p}
	\fmfiset{v3}{point 0 of p}
	\fmfi{#2}{subpath (0,-length(p)/4) of p}
	\fmfi{#1}{subpath (0,3length(p)/4) of p}
	\fmfi{#3}{v1--v3}
	\fmfi{#4}{v2--v1}
  \end{fmfgraph}}}

\def\Basketball#1#2#3#4{%
  \parbox{45\unitlength}{
  \begin{fmfgraph}(45,30)
	\fmfipath{p[]}
	\fmfiset{p1}{fullcircle scaled 30 rotated -60 shifted (15,15) }
	\fmfiset{p2}{fullcircle scaled 30 rotated -120 shifted (30,15) }
	\fmfi{#1}{subpath (length(p1)/3,length(p1)) of p1}
	\fmfi{#2}{subpath (2length(p2)/3,length(p2)) of p2}
	\fmfi{#3}{subpath (0,length(p1)/3) of p1}
	\fmfi{#4}{subpath (0,2length(p2)/3) of p2}
  \end{fmfgraph}}}
  
  
\def\RingRing#1#2#3{%
  \parbox{60\unitlength}{
  \begin{fmfgraph}(60,40)
	\fmfi{#2}{fullcircle scaled .5h shifted (w/6,0.5h)}
	\fmfi{#3}{fullcircle scaled .5h shifted (5w/6,0.5h)}
	\fmfipath{p}
	\fmfiset{p}{fullcircle scaled 1h shifted (.5w,.5h)}
	\fmfi{#1}{subpath (2*angle(sqrt(15),1)*length(p)/360,(1/2-2*angle(sqrt(15),1)/360)*length(p)) of p}
	\fmfi{#1}{subpath ((1/2+2*angle(sqrt(15),1)/360)*length(p),(1-2*angle(sqrt(15),1)/360)*length(p)) of p}
  \end{fmfgraph}}}

\def\FlatFlat#1#2#3{%
  \parbox{60\unitlength}{
  \begin{fmfgraph}(60,40)
	\fmfipath{p[]}
	\fmfi{#1}{halfcircle scaled 1h shifted (.5w,.5h)}
	\fmfi{#1}{halfcircle scaled 1h rotated 180 shifted (.5w,.5h)}
	\fmfiset{p2}{fullcircle scaled .5h shifted (w/6,0.5h)}
	\fmfiset{p3}{fullcircle scaled .5h rotated 180 shifted (5w/6,0.5h)}
	\fmfi{#2}{subpath (angle(1,sqrt(15))*length(p2)/360,(1-angle(1,sqrt(15))/360)*length(p2)) of p2}
	\fmfi{#3}{subpath (angle(1,sqrt(15))*length(p3)/360,(1-angle(1,sqrt(15))/360)*length(p3)) of p3}
  \end{fmfgraph}}}

\def\LoopLoop#1#2#3{%
  \parbox{90\unitlength}{
  \begin{fmfgraph}(90,30)
	\fmfi{#1}{halfcircle scaled 1h shifted (.5w,.5h)}
	\fmfi{#1}{halfcircle scaled 1h rotated 180 shifted (.5w,.5h)}
	\fmfi{#2}{fullcircle scaled 1h shifted (w/6,.5h)}
	\fmfi{#3}{fullcircle scaled 1h rotated 180 shifted (5w/6,.5h)}
  \end{fmfgraph}}}

\def\RingLoop#1#2#3{%
  \parbox{70\unitlength}{
  \begin{fmfgraph}(70,40)
	\fmfi{#2}{fullcircle scaled .5h shifted (w/7,0.5h)}
	\fmfi{#3}{fullcircle scaled .5h rotated 180 shifted (6w/7,0.5h)}
	\fmfipath{p}
	\fmfiset{p}{fullcircle scaled 1h shifted (3w/7,.5h)}
	\fmfi{#1}{subpath (0,(1/2-2*angle(sqrt(15),1)/360)*length(p)) of p}
	\fmfi{#1}{subpath ((1/2+2*angle(sqrt(15),1)/360)*length(p),length(p)) of p}
  \end{fmfgraph}}}

\def\FlatLoop#1#2#3{%
  \parbox{70\unitlength}{
  \begin{fmfgraph}(70,40)
	\fmfi{#1}{halfcircle scaled 1h shifted (3w/7,.5h)}
	\fmfi{#1}{halfcircle scaled 1h rotated 180 shifted (3w/7,.5h)}
	\fmfipath{p}
	\fmfiset{p}{fullcircle scaled .5h shifted (w/7,0.5h)}
	\fmfi{#2}{subpath (angle(1,sqrt(15))*length(p)/360,(1-angle(1,sqrt(15))/360)*length(p)) of p}
	\fmfi{#3}{fullcircle scaled .5h rotated 180 shifted (6w/7,0.5h)}
  \end{fmfgraph}}}

\def\RingFlat#1#2#3{%
  \parbox{60\unitlength}{
  \begin{fmfgraph}(60,40)
	\fmfipath{p[]}
	\fmfiset{p}{fullcircle scaled 1h shifted (.5w,.5h)} 
	\fmfi{#1}{subpath (0,(1/2-2*angle(sqrt(15),1)/360)*length(p)) of p}
	\fmfi{#1}{subpath ((1/2+2*angle(sqrt(15),1)/360)*length(p),length(p)) of p}
	\fmfiset{p3}{fullcircle scaled .5h rotated 180 shifted (5w/6,0.5h)}
	\fmfi{#2}{fullcircle scaled .5h shifted (w/6,0.5h)}
	\fmfi{#3}{subpath (angle(1,sqrt(15))*length(p3)/360,(1-angle(1,sqrt(15))/360)*length(p3)) of p3}
  \end{fmfgraph}}}

  
\def\SelfenA#1#2{%
  \begin{fmfgraph}(60,30)
	\fmfipair{v[]}
	\fmfi{#2}{halfcircle scaled 1h shifted (0.5w,0.5h)}
	\fmfi{#2}{halfcircle scaled 1h rotated 180 shifted (0.5w,0.5h)}
	\fmfiset{v1}{(0,0.5h)}
	\fmfiset{v2}{(w,0.5h)}
	\fmfi{#1}{v1--(0.5w-0.5h,0.5h)}
	\fmfi{#1}{(0.5w+0.5h,0.5h)--v2}
  \end{fmfgraph}}

\def\SelfenAflat#1#2#3{%
  \begin{fmfgraph}(60,30)
	\fmfi{#2}{halfcircle scaled 1h shifted (0.5w,0)}
	\fmfi{#3}{(0,0)--(0.5w-0.5h,0)}
	\fmfi{#1}{(0.5w-0.5h,0)--(0.5w+0.5h,0)}
	\fmfi{#3}{(0.5w+0.5h,0)--(1w,0)}
  \end{fmfgraph}}
	
\def\SelfenB#1#2{%
  \begin{fmfgraph}(50,30)
	\fmfipair{v}
	\fmfi{#2}{fullcircle scaled 1h rotated -90 shifted (0.5w,0.5h)}
	\fmfiset{v}{(0.5w,0)}
	\fmfi{#1}{(0,0)--v}
	\fmfi{#1}{v--(w,0)}
  \end{fmfgraph}}


\fmfcmd{%
  style_def dash_dot expr p =
    save dpp, k;
    numeric dpp, k;
    dpp = ceiling (pixlen (p, 10) / (1.5*dash_len)) / length p;
    k=0;
    forever:
      exitif k+.33 > dpp*length(p);
      cdraw point k/dpp of p .. point (k+.33)/dpp of p;
      exitif k+.67 > dpp*length(p);
      cdrawdot point (k+.67)/dpp  of p;
      k := k+1;
    endfor
  enddef;}
  
\fmfcmd{%
  style_def dash_dot_arrow expr p =
    draw_dash_dot p;
    cfill (arrow p);
  enddef;}

\fmfset{arrow_len}{2.5mm}
\fmfset{dash_len}{2.5mm}

\begin{eqnarray}
-2 \times \Ring{fermion} & = & -\left[1 + 2\Nc + \dF\left(1 + \Nc\right)\right]\NF\widetilde{\Iint}_0' \\
2 \times \Ring{scalar} & = & \dF\nS \left[\Iint_0' + \nu^2 \Iint_1 + \frac{1}{2}\nu^4 \Iint_2\right] + \mathcal{O}\left(\nu^6\right) \\
1\times \Ring{photon, w=2} & = & \frac{1}{2}D\left(\dA + 1\right)\Iint_0' \\
-2 \times \Ring{ghost, w=2} & = & -\left(\dA + 1\right)\Iint_0'
\end{eqnarray}

\begin{eqnarray}
\frac{1}{8} \times \DiaEight{photon}{photon} & = & -\frac{1}{4}D\left(D-1\right)\CA\dA\Z_g g^2\Iint_1^2 \\
\frac{1}{2} \times \DiaEight{scalar}{scalar} & = & -\dF\left(1 + \dF\right)\Z_\lambda \lambda \nS\left(\Iint_1^2 + 2\nu^2\Iint_1\Iint_2\right) \\
\frac{1}{2} \times \DiaEight{scalar}{photon, w=2} & = & -\frac{1}{4}D\dF\left(4\CF\Z_g g^2 + \Z_{g'} g'^2\right)\nS\left(\Iint_1^2 + \nu^2\Iint_1\Iint_2\right) \\
\frac{1}{12} \times \Sunset{photon}{photon} & = & \frac{3}{4}\left(D-1\right)\CA\dA\Z_g g^2\Iint_1^2 \\
-\frac{1}{2} \times \Sunset{ghost}{photon} & = & -\frac{1}{4}\CA\dA\Z_g g^2 \Iint_1^2 \\
-\frac{1}{2} \times \Sunset{fermion, w=2}{photon, w=2} & = & \frac{2-D}{2}\left[\CF\dF\left(1+\Nc\right)\Z_g g^2 + \left(1+\frac{\dF}{4} + \frac{20+\dF}{36}\Nc\right)\Z_{g'} g'^2\right]\NF \nonumber \\ 
& & \times \left(\widetilde{\Iint}_1^2 - 2\widetilde{\Iint}_1\Iint_1\right) \\
\frac{1}{2} \times \Sunset{scalar}{photon, w=2} & = & \frac{1}{2}\dF\left(\CF\Z_g g^2 + \frac{1}{4}\Z_{g'}g'^2\right)\nS\left(3\Iint_1^2 + 2\nu^2\Iint_1\Iint_2\right) \\
-1\times \FSunset{fermion, w=2}{fermion}{scalar} & = & 2\Nc\Z_Y^2g_Y^2\left(2\Iint_1\widetilde{\Iint}_1 - \widetilde{\Iint}_1^2 + 2\nu^2\widetilde{\Iint}_1\Iint_2\right)
\end{eqnarray}

\begin{eqnarray}
\frac{1}{24} \times \Mersu{photon}{photon}{photon}{photon} & = & \frac{1}{8}\left(5D-5-\frac{3}{4}\right)\CA^2\dA g^4\Mint_{0,0} \\
-\frac{1}{3} \times \Mersu{ghost}{ghost}{photon}{photon} & = & -\frac{1}{16}\CA^2\dA g^4\Mint_{0,0} \\
-\frac{1}{4} \times \Mersu{ghost}{photon}{photon}{ghost} & = & -\frac{1}{32}\CA^2\dA g^4\Mint_{0,0} \\
-\frac{1}{3} \times \Mersu{fermion}{fermion}{photon}{photon} & = & \frac{1}{2}\left(D-2\right)\CA\CF\dF\NF\left(1+\Nc\right)g^4 \widetilde{\Mint}_{0,0} \\
-\frac{1}{4} \times \Mersu{fermion, w=2}{photon, w=2}{photon, w=2}{fermion, w=2} & = & \frac{1}{8}\left(D-2\right)
\left[\left(\CA\CF\dF-2\CF^2\dF\right)\left(1+\Nc\right)g^4 - \CF\dF\left(1+\frac{\Nc}{9}\right)g^2g'^2 \right. \nonumber \\
& & \left. \hspace{1.8cm} - \frac{1}{8}\left(16+\dF+\frac{272+\dF}{81}\Nc\right)g'^4\right]\NF \nonumber \\
& & \times \left[2\left(4-D\right)\widetilde{\Mint}_{0,0} + \left(D-6\right)\Nint_{0,0}\right] \\
\frac{1}{3} \times \Mersu{scalar}{scalar}{photon}{photon} & = & \frac{5}{8}\CA\CF\dF g^4 \nS \Mint_{0,0} \\
\frac{1}{4} \times \Mersu{scalar}{photon, w=2}{photon, w=2}{scalar} & = & \frac{5}{8} \left(-\left(\CA\CF\dF-2\CF^2\dF\right)g^4 + \CF\dF g^2g'^2 + \frac{\dF}{8}g'^4\right)\nS\Mint_{0,0} \\
-\frac{1}{2} \times \Mersu{fermion, w=2}{photon, w=2}{gluon}{fermion, w=2} & = & -\frac{1}{4}\left(D-2\right)\left(\CF\dF g^2g_s^2 + \frac{20+\dF}{36}g'^2g_s^2\right)\NF\left(\Nc^2-1\right) \nonumber \\
& & \times \left[2\left(4-D\right)\widetilde{\Mint}_{0,0} + \left(D-6\right)\Nint_{0,0}\right] \\
-1\times \Mersu{fermion, w=2}{dashes}{photon, w=2}{fermion, w=2} & = & \frac{8}{9}\frac{\CF}{\dA}\Nc g'^2g_Y^2\left[\left(2D-4\right)\widetilde{\Mint}_{0,0} + \left(3-D\right)\Nint_{0,0}\right] \\
-1\times \FMersu{dashes}{fermion, w=2}{photon, w=2}{fermion, w=2} & = & \frac{1}{2}\Nc\left(3g^2g_Y^2 + g'^2g_Y^2\right)\widetilde{\Mint}_{0,0} \\
-1\times \Mersu{fermion, w=2}{dashes}{gluon}{fermion, w=2} & = & \left(\Nc^2 - 1\right)g_Y^2g_s^2\left[\left(2D-4\right)\widetilde{\Mint}_{0,0} + \left(3-D\right)\Nint_{0,0}\right] \\
\frac{1}{48} \times \Basketball{photon}{photon}{photon}{photon} & = & \frac{3}{16}D\left(D-1\right)\CA^2\dA g^4\Mint_{0,0}\\
\frac{1}{4} \times \Basketball{scalar}{photon, w=2}{scalar}{photon, w=2} & = & \frac{1}{4}D\left[\left(4\CF^2\dF-\CA\CF\dF\right) g^4 + 2\CF\dF g^2g'^2 + \frac{1}{4}\dF g'^4\right]\nS\Mint_{0,0} \\
\frac{1}{8} \times \Basketball{scalar}{scalar}{scalar}{scalar} & = & \dF\left(1+\dF\right)\lambda^2\nS\Mint_{0,0} \\
\frac{1}{8} \times \FDiaV{photon}{photon}{photon}{photon} & = & -\frac{27}{16}\left(D-1\right)\CA^2\dA g^4\Mint_{0,0} \\
1\times \FDiaV{photon, w=2}{scalar}{scalar}{scalar} & = & -\frac{9}{8}\left[\left(4\CF^2\dF - \CA\CF\dF\right) g^4 + 2\CF\dF g^2g'^2 + \frac{\dF}{4} g'^4\right]\nS\Mint_{0,0} \\
\frac{1}{2} \times \DiaV{scalar}{photon, w=2}{scalar}{scalar} & = & 0
\end{eqnarray}

\begin{eqnarray}
  \frac{1}{16} & \times & \left[\RingRing{photon}{photon}{photon} 
\; + 2 \times \; \RingLoop{photon}{photon}{photon}
\; + \; \LoopLoop{photon}{photon}{photon} \right. \nonumber \\
& - &\left. 4 \times \RingRing{photon}{ghost}{photon}
\; - 4 \times \RingLoop{photon}{ghost}{photon} 
\; + 4 \times \RingRing{photon}{ghost}{ghost}\right] \nonumber \\
& = & \frac{1}{4}\CA^2\dA g^4\left[ \left(D-2\right)^2\Mint_{2,-2} - \left(\left(\frac{D+2}{2}\right)^2 -4D\right)\Mint_{0,0} + \left(D-6\right)\left(D-2\right)^2\Iint_1^2\Iint_2\right] \\
-\frac{1}{4} & \times & \left[\RingRing{photon}{fermion}{photon}
\; - 2 \times \RingRing{photon}{fermion}{ghost}
\; + \RingLoop{photon}{fermion}{photon}\right] \nonumber \\
& = & -\half \CA\CF\dF\NF\left(1+\Nc\right)g^4(D-2)\left[2\widetilde{\Mint}_{-2,2} + \widetilde{\Mint}_{0,0} + 2\left(D-6\right)\Iint_1\widetilde{\Iint}_1\Iint_2\right] \\
\frac{1}{4} & \times & \RingRing{photon, w=2}{fermion, w=2}{fermion, w=2} \;\; = \frac{1}{4}\NF^2\left\{\CF\TF\dF\left(1+\Nc\right)^2g^4 + \left[1+\frac{\dF}{4} + \frac{\Nc}{36}\left(20+\dF\right)\right]^2g'^4\right\} \nonumber \\
& & \times \left[4\Nint_{2,-2} + \left(D-4\right)\Nint_{0,0} - 4(6-D)\widetilde{\Iint}_1^2\Iint_2\right] \\
\frac{1}{4} & \times & \left[\RingRing{photon}{scalar}{photon}
\; - 2 \times \RingRing{photon}{scalar}{ghost}
\; + \RingLoop{photon}{scalar}{photon}\right] \nonumber \\
& = & \frac{1}{2}\CA\CF\dF\nS g^4\left[2(D-2)\Mint_{2,-2}-\frac{D+2}{2}\Mint_{0,0}-8(D-2)\Iint_1^2\Iint_2\right] \\
\frac{1}{4} & \times & \left[\RingLoop{photon}{photon}{scalar}
\; - 2 \times \RingLoop{photon}{ghost}{scalar}
\; + \LoopLoop{photon}{photon}{scalar}\right] \nonumber \\
& = & \CA\CF\dF(D-2)^2 \nS g^4 \Iint_1^2\Iint_2 \\
\frac{1}{4} & \times & \left[\RingRing{photon, w=2}{scalar}{scalar} + 2\times \RingLoop{photon, w=2}{scalar}{scalar} + \LoopLoop{photon, w=2}{scalar}{scalar}\right] \nonumber \\
 & = & \frac{1}{4}\left(\CF\TF\dF g^4 + \frac{\dF^2}{16}g'^4\right)\nS\left[4\Mint_{2,-2} - \Mint_{0,0} +
  4\left(D-6\right)\Iint_1^2\Iint_2\right] \\
-\frac{1}{2} & \times & \left[\RingRing{photon, w=2}{fermion, w=2}{scalar} + \RingLoop{photon, w=2}{fermion, w=2}{scalar}\right] \nonumber \\
& = & -\frac{1}{2}\left[4\CF\TF\dF(1+\Nc)g^4 
  + \dF\left(1+\frac{\dF}{4} + \frac{20+\dF}{36}\Nc\right)g'^4\right]\NF\nS \nonumber \\
& & \hspace{1.1cm}\times \left[\widetilde{\Mint}_{-2,2} - \frac{1}{2}\widetilde{\Mint}_{0,0} + \left(D-6\right)\Iint_1\widetilde{\Iint}_2\Iint_2\right]
\end{eqnarray}
\begin{eqnarray}
-\frac{1}{2} \times \FlatFlat{ghost}{photon}{photon} & = & -\frac{1}{8}\CA^2\dA g^4\Mint_{0,0} \\
-\frac{1}{2} \times \FlatFlat{fermion, w=2}{photon, w=2}{photon, w=2} & = & -\NF\left(2-D\right)^2\left\{\frac{1}{2}\CF^2\dF\left(1+\Nc\right)g^4 
+ \left[\left(\half+\frac{17}{162}\Nc\right) + \frac{\dF}{32}\left(1+\frac{\Nc}{81}\right)\right]g'^4 \right. \nonumber \\ 
& & \left. + \frac{1}{4}\CF\dF\left(1+\frac{\Nc}{9}\right)g^2g'^2\right\} \times \left[\widetilde{\Mint}_{1,-1} + \widetilde{\Mint}_{0,0} + \left(\Iint_1 - \widetilde{\Iint}_1\right)^2\widetilde{\Iint}_2\right] \\
-1\times \;\; \FlatFlat{fermion, w=2}{gluon}{photon, w=2} & = & -\frac{1}{2}\left(D-2\right)^2\NF\left(\Nc^2-1\right)
   \left(\CF\dF g^2g_s^2 + \frac{\dF+20}{36}g'^2g_s^2\right) \nonumber \\ 
& & \times \left[\widetilde{\Mint}_{1,-1} + \widetilde{\Mint}_{0,0} + \left(\Iint_1 - \widetilde{\Iint}_1\right)^2\widetilde{\Iint}_2\right] \\
-\frac{1}{2} \times \FlatFlat{fermion, w=2}{dashes}{dashes} & = & -3\Nc g_Y^4\left[\widetilde{\Mint}_{1,-1} + \widetilde{\Mint}_{0,0} + \left(\Iint_1 - \widetilde{\Iint}_1\right)^2\widetilde{\Iint}_2\right] \\
-1\times \FlatFlat{fermion, w=2}{dashes}{photon, w=2} & = & -\left(D-2\right)\Nc\left(\frac{3}{2} g^2g_Y^2 + \frac{17}{18} g'^2g_Y^2\right) \nonumber \\
& & \times \left[\widetilde{\Mint}_{1,-1} + \widetilde{\Mint}_{0,0} + \left(\Iint_1 - \widetilde{\Iint}_1\right)^2\widetilde{\Iint}_2\right]  \\
-1\times \FlatFlat{fermion, w=2}{dashes}{gluon} & = & -\left(2D-4\right)\left(\Nc^2-1\right)g_Y^2g_s^2\left[\widetilde{\Mint}_{1,-1} + \widetilde{\Mint}_{0,0} + \left(\Iint_1-\widetilde{\Iint}_1\right)^2\widetilde{\Iint}_2\right]
\end{eqnarray}
\begin{eqnarray}
\frac{1}{2} \times \LoopLoop{scalar}{scalar}{scalar} & = & 2\dF\left(\dF+1\right)^2\nS\lambda^2\Iint_1^2\Iint_2 \\
\frac{1}{2} \times \LoopLoop{scalar}{photon, w=2}{scalar} & = & \frac{1}{2}D\dF(\dF+1)\left(4\CF g^2\lambda + g'^2\lambda\right)\nS\Iint_1^2\Iint_2 \\
1\times \FlatLoop{scalar}{photon, w=2}{scalar} & = & -\frac{1}{2}\dF(\dF+1)\left(4\CF g^2\lambda + g'^2\lambda\right)\nS\Iint_1^2\Iint_2 \\
-1\times \RingLoop{scalar}{fermion, w=2}{scalar} & = & -24 \Nc \lambda g_Y^2 \Iint_1\widetilde{\Iint}_1\Iint_2 \\
\frac{1}{2} \times \FlatLoop{scalar}{photon, w=2}{photon, w=2} & = & -D\left[\CF^2\dF g^4 + \frac{1}{16}\dF g'^4 + \frac{1}{2}\CF\dF g^2g'^2\right]\nS\Iint_1^2\Iint_2 \\
\frac{1}{2} \times \FlatFlat{scalar}{photon, w=2}{photon, w=2} & = & \left(\frac{1}{2}\CF^2\dF g^4 + \frac{1}{32}\dF g'^4 + \frac{1}{4}\CF\dF g^2g'^2\right)\nS\left(4\Mint_{0,0} + \Iint_1^2\Iint_2\right)  \\
\frac{1}{8} \times \LoopLoop{scalar}{photon, w=2}{photon, w=2} & = & D^2\left(\frac{1}{2}\CF^2\dF g^4 + \frac{1}{32}\dF g'^4 + \frac{1}{4}\CF\dF g^2g'^2\right)\nS\Iint_1^2\Iint_2 \\ 
\frac{1}{2} \times \RingRing{scalar}{fermion, w=2}{fermion, w=2} & = & \Nc^2g_Y^4\left(4\widetilde{\Iint}_1^2\Iint_2 + \Nint_{0,0}\right) \\
-\frac{1}{2} \times \RingLoop{scalar}{fermion, w=2}{photon, w=2} & = & -D\Nc\left( 3 g^2g_Y^2 + g'^2g_Y^2\right)\Iint_1\widetilde{\Iint}_1\Iint_2 \\
-1\times \RingFlat{scalar}{fermion, w=2}{photon, w=2} & = & -\Nc\left(3g^2 g_Y^2 + g'^2g_Y^2\right)\left(\widetilde{\Mint}_{0,0} - \Iint_1\widetilde{\Iint}_1\Iint_2 \right)
\end{eqnarray}
\end{fmffile}


\subsection{Diagrams in the effective theory}
\label{et_diagrams}

The required $d=3-2\epsilon$ dimensional integrals have been worked out in the literature except for the particular combination of diffenent masses and massless propagators in Eq. (\ref{eq:hard_integral_dia}). The results are given here in terms of the integrals $I$, $J$, $K$, $M$, $N$ and $L$ given in the 

\begin{fmffile}{et_diags}

\fmfset{arrow_len}{2.5mm}
\fmfset{dash_len}{2.5mm}

\begin{eqnarray}
	2 \times \Ring{scalar} &=& \dF I'_0(m_3)
\\
	1\times \Ring{dash_dot,w=2} &=& \half\dA I'_0(\mD)+\half I'_0(\mD')
\\
	\half \times \DiaEight{scalar}{scalar} &=& -\dF(\dF+1) \lambda_3 I_1^2(m_3)
\\
	\half \times \DiaEight{scalar}{dash_dot,w=2} &=& -\dF \dA h_3 I_1(m_3) I_1(\mD) -\dF h_3' I_1(m_3) I_1(\mD')
\\
	\half \times \Sunset{scalar}{photon,w=2} 
	&=& -\half\dF\left(\CF g_3^2 +\frac{1}{4}g_3'^2\right) \left[ I_1^2(m_3)+4m_3^2J_1(m_3) \right]
\\
	\frac{1}{4} \times \Sunset{dash_dot}{photon}
	&=& -\frac{1}{4} \CA \dA g_3^2 \left[ I_1^2(\mD)+4\mD^2J_1(\mD) \right]
\end{eqnarray}

\begin{eqnarray}
	\frac{1}{3} \times \Mersu{scalar}{scalar}{photon}{photon} 	
	&=& \CA\CF\dF g_3^4 \left[ -\frac{1}{4}M_{1,-1} -M_{-1,1} -\half I_1 J_1 -2 m_3^2 M_{1,0} \right]_{m=m_3}
\\
	\frac{1}{4} \times \Mersu{scalar}{photon,w=2}{photon,w=2}{scalar}	
	& = & \left[ \left( \CF^2\dF -\half\CA\CF\dF \right) g_3^4 +2\CF\dF\frac{1}{4} g_3^2 g_3'^2 +\frac{\dF}{16}g_3'^4\right] \nonumber \\ 
	&&\times\left[ N_{1,-1}(m_3)+L_{1,-1}(m_3) +M_{0,0}(m_3) -6I_1(m_3)J_1(m_3)+\frac{5}{4}N_{0,0}(m_3) \right. \nonumber \\ 
	&& \left. {}+6m_3^2N_{1,0}(m_3) -8m_3^2M_{1,0}(m_3)+4m_3^4N_{1,1}(m_3) \right] 
\\
	\frac{1}{6} \times \Mersu{dash_dot}{dash_dot}{photon}{photon}	
	& = & \CA^2\dA g_3^4 \left[-\frac{1}{8}M_{1,-1} -\half M_{-1,1} -\frac{1}{4}I_1 J_1 -\mD^2 M_{1,0} \right]_{m=\mD}
\\
	\frac{1}{8} \times \Mersu{dash_dot}{photon}{photon}{dash_dot}	
	& = & \CA^2\dA g_3^4 \left[ \frac{1}{4}N_{1,-1}(\mD)+\frac{1}{4}L_{1,-1}(\mD) +\frac{1}{4} M_{0,0}(\mD)
	-\frac{3}{2}I_1(\mD)J_1(\mD) \right. \nonumber \\
	&& \left. {}+\frac{5}{16}N_{0,0}(\mD) +\frac{3}{2}\mD^2N_{1,0}(\mD) -2\mD^2 M_{1,0}(\mD) +\mD^4 N_{1,1}(\mD) \right]
\end{eqnarray}

\begin{eqnarray}
	1 \times \DiaV{scalar}{scalar}{photon,w=2}{photon,w=2}	
	& = &  \left[\left(\CF^2\dF-\frac{1}{4}\CA\CF\dF\right)g_3^4 +2 \CF\dF\frac{1}{4}g_3^2g_3'^2 +\frac{\dF}{16}g_3'^4\right] \nonumber \\
	&& \times \left[ M_{1,-1}(m_3) -4 M_{0,0}(m_3)+2I_1(m_3)J_1(m_3) +8m_3^2 M_{1,0}(m_3) \right]
\\
	\half \times \DiaV{scalar}{photon,w=2}{scalar}{scalar}	
	& = & \left( \CF\dF g_3^2\lambda_3+\frac{\dF(\dF+1)}{4}g_3'^2\lambda_3\right)
	 \left[ N_{0,0} -4 I_1 J_1 +4 m_3^2 N_{1,0} \right]_{m=m_3}
\\
	\half \times \DiaV{dash_dot}{dash_dot}{photon}{photon}
	& = & \CA^2 \dA g_3^4 \left[ \frac{3}{8} M_{1,-1} -\frac{3}{2}M_{0,0} +\frac{3}{4}I_1 J_1
	+3\mD^2M_{1,0} \right]_{m=\mD}
\end{eqnarray}

\begin{eqnarray}
	\frac{1}{4} \times \Basketball{scalar}{photon,w=2}{scalar}{photon,w=2}
	& = & \left[\left(\CF^2\dF-\frac{1}{4}\CA\CF\dF\right)g_3^4 +2 \CF\dF\frac{1}{4}g_3^2g_3'^2 +\frac{\dF}{16}g_3'^4\right] d\; M_{0,0}(m_3)
\\
	\frac{1}{8} \times \Basketball{scalar}{scalar}{scalar}{scalar}
	& = & \dF (\dF+1) \lambda_3^2 N_{0,0}(m_3)
\\
	\frac{1}{8} \times \Basketball{dash_dot}{photon}{dash_dot}{photon}
	& = & \CA^2\dA g_3^4 \frac{3d}{8}M_{0,0}(\mD)
\\
	\frac{1}{4} \times \Basketball{scalar}{dash_dot,w=2}{scalar}{dash_dot,w=2}
	& = &  h_3^2\dF\dA \AIE(m_3,m_3,\mD,\mD) +h_3'^2\dF\AIE(m_3,m_3,\mD',\mD') \nonumber \\
	&&{}+2\CF\dF\frac{1}{4} g_3^2 g_3'^2\AIE(m_3,m_3,\mD,\mD')
\end{eqnarray}

\begin{eqnarray}
\frac{1}{4} &\times & \left[
	\RingRing{photon}{photon}{scalar} +
	\RingLoop{photon}{photon}{scalar} +
	\RingRing{photon}{ghost}{scalar} +
	\RingLoop{photon}{ghost}{scalar} \right] \nonumber \\
	&=& -\CA\CF\dF g_3^4\frac{3d-2}{4(d-1)}\left[ M_{0,0} +4m_3^2 M_{0,1}\right]_{m=m_3} \\
\frac{1}{4} &\times & \left[
	\RingRing{photon,w=2}{scalar}{scalar} +
	\LoopLoop{photon,w=2}{scalar}{scalar} +
	\RingLoop{photon,w=2}{scalar}{scalar} \right]
	= \left(\CF\TF\dF g_3^4 + \frac{\dF^2}{16}g_3'^4 \right) \nonumber \\
	&&\times \left[ \frac{d-2}{d-1}I_1\left(J_1+4m_3^2 K_2\right) +\frac{1}{4(d-1)}\left(N_{0,0} +8m_3^2N_{1,0} +16m_3^4 N_{2,0}\right)\right]_{m=m_3} \\
\frac{1}{4} &\times & \left[
	\RingRing{photon}{photon}{dash_dot} +
	\RingLoop{photon}{photon}{dash_dot} +
	\RingRing{photon}{ghost}{dash_dot} +
	\RingLoop{photon}{ghost}{dash_dot} \right] \nonumber \\
	&=& -\CA^2\dA g_3^4\frac{3d-2}{8(d-1)}\left[ M_{0,0} +4\mD^2 M_{0,1}\right]_{m=\mD} \\
\frac{1}{4} &\times & \left[
	\RingRing{photon}{dash_dot}{dash_dot} +
	\LoopLoop{photon}{dash_dot}{dash_dot} +
	\RingLoop{photon}{dash_dot}{dash_dot} \right] \nonumber \\
	&=& \CA^2\dA g_3^4 \left[ \frac{d-2}{4(d-1)}I_1\left(J_1+4\mD^2 K_2\right) +\frac{1}{16(d-1)}\left(N_{0,0} +8\mD^2N_{1,0} +16\mD^4 N_{2,0}\right)\right]_{m=\mD} \\
\frac{1}{4} &\times & \left[
	\RingRing{photon}{scalar}{dash_dot} +
	\RingLoop{photon}{scalar}{dash_dot} +
	\RingLoop{photon}{dash_dot}{scalar} \right] \nonumber \\	\label{eq:hard_integral_dia}
	&=& \CA\CF\dF g_3^4 \left\{ \frac{1}{4(d-1)}\AIE(m_3,m_3,\mD,\mD) +\frac{1}{d-1}(m_3^2+\mD^2)\AIC(m_3,m_3,\mD,\mD,0)  \right. \nonumber \\
	&& {}+\frac{d-2}{2(d-1)}\left[ I_1(m_3)\left(J_1(\mD)+4\mD^2K_2(\mD)\right) +I_1(\mD) \left(J_1(m_3) +4m_3^2K_2(m_3)\right)\right] \nonumber \\ 
	&& \left. {}+\frac{1}{(4\pi)^3}\frac{1}{6(d-1)}\left[ \frac{\mD^2}{m_3}\ln\frac{m_3+\mD}{\mD} +\frac{m_3^2}{\mD}\ln\frac{m_3+\mD}{m_3} -4(m_3+\mD)\right] \right\}
\end{eqnarray}

\begin{eqnarray}
\half \times 
	\FlatFlat{scalar}{photon,w=2}{photon,w=2} 
	& = & \left( \CF^2\dF g_3^4 +2\CF\dF\frac{1}{4}g_3^2 g_3'^2 +\frac{\dF}{16}g_3'^4 \right) 
	\left[ \half I_1^2 I_2 -2I_1 J_1 +2M_{0,0} \right. \nonumber \\ 
	&& \left. {}+4m_3^2I_1 J_2 -8m_3^2 M_{1,0} +8m_3^4M_{2,0} \right]_{m=m_3} \\
1 \times \FlatLoop{scalar}{photon,w=2}{scalar}
	& = & \left(\CF g_3^2 +\frac{1}{4}g_3'^2\right)\dF(\dF+1)\lambda_3 \left[ 2I_1^2 I_2 -4I_1J_1 +8m_3^2I_1J_2 \right]_{m=m_3} \\
\half \times \FlatLoop{scalar}{photon,w=2}{dash_dot,w=2}
	& = & \left(\CF\dF g_3^2 +\frac{1}{4}\dF g_3'^2\right) \left( \dA h_3 I_1(\mD) +h_3' I_1(\mD')\right) \nonumber \\
	&& \times\left[ I_1(m_3) I_2(m_3) -2J_1(m_3) +4m_3^2 J_2(m_3) \right]
\end{eqnarray}

\begin{eqnarray}
\half \times \LoopLoop{scalar}{scalar}{scalar}
	&=& \dF(\dF+1)^2 \lambda_3^2 I_1^2(m_3) I_2(m_3) \\
\frac{1}{8} \times \LoopLoop{scalar}{dash_dot,w=2}{dash_dot,w=2}
	&=& \half \dF \left[ \dA h_3 I_1(\mD) +h_3' I_1(\mD')\right]^2 I_2(m_3) \\
\half \times \LoopLoop{scalar}{scalar}{dash_dot,w=2}
	&=& \left( \dA h_3 I_1(\mD) +h_3' I_1(\mD')\right)\dF(\dF+1)\lambda_3 2I_1(m_3)I_2(m_3)
\end{eqnarray}

\begin{eqnarray}
\frac{1}{4} &\times & \FlatFlat{dash_dot}{photon}{photon} 
	= \CA^2\dA g_3^4 \left[ \frac{1}{4} I_1^2I_2 -I_1 J_1 +M_{0,0} +2\mD^2I_1 J_2 -4\mD^2 M_{1,0} +4\mD^4M_{2,0} \right]_{m=\mD} \\
\half &\times  &\FlatLoop{dash_dot}{photon}{scalar}
	= \CA\dA\dF g_3^2 h_3 I_1(m_3)\left[ I_1(\mD) I_2(\mD) -2J_1(\mD) +4\mD^2 J_2(\mD) \right] \\
\frac{1}{4} &\times & \LoopLoop{dash_dot,w=2}{scalar}{scalar}
	= \left( \dA h_3^2 I_2(\mD) +h'^2_3 I_2(\mD')\right)\dF^2 I_1^2(m_3)
\end{eqnarray}

\end{fmffile}

\bibliography{articles}

\end{document}